\documentclass[english,reprint,superscriptaddress,showpacs,preprintnumbers,nofootinbib,amsmath,amssymb,aps,onecolumn,prd,
floatfix]{revtex4-1}
\usepackage{array}
\usepackage{graphicx}
\usepackage{epsfig}
\usepackage{dcolumn}
\usepackage{bm}
\usepackage{amssymb}
\usepackage{amsmath}
\usepackage{subfigure}
\usepackage{cancel}
\usepackage[colorlinks]{hyperref}
\usepackage[usenames,dvipsnames]{color}

\hypersetup{
     breaklinks=true,
    pdfstartview={FitH},    
    colorlinks=true,       
    linkcolor=blue,          
    citecolor=red,        
    filecolor=magenta,      
    urlcolor=blue,           
    anchorcolor=green,      
    linktocpage=true
}

\begin{document}

\renewcommand{\theequation}{\arabic{section}.\arabic{equation}}
\renewcommand{\thefigure}{\arabic{section}.\arabic{figure}}
\renewcommand{\thetable}{\arabic{section}.\arabic{table}}

\title{\Large \baselineskip=16pt  
Power Law Plateau Inflation Potential In The RS II Braneworld Evading Swampland Conjecture }

\author{$Rathin~Adhikari^{1}, Mayukh~R.~Gangopadhyay^{1}$,$Yogesh$}
\affiliation{
\it Centre For Theoretical Physics, Jamia Millia Islamia, New Delhi-110025, India.}
\begin{abstract}\normalsize \baselineskip=10pt 
In the recent time,  inflationary cosmology is facing an existential crisis due to the proposed Swampland criterion which aims to evade any (meta-)stable de Sitter construction within the String landscape.  It is been realised that a single field slow roll inflation is inconsistent with the Swampland criterion unless the inflationary model is realised in some non-standard scenario such as Warm inflation or the Braneworld scenario. In \cite{Dimopoulos}, Dimopoulos and Owen introduced a new class of model of inflation dubbed as the power law plateau inflation in the standard cold inflationary scenario. But to realise this model in the standard scenario consistent with observation, they had to introduce a phase of thermal inflation. In this paper we have analysed this model in the braneworld scenario to show that for some choice of the parameters defining the model class, one can have an observationally consistent power law plateau without any phase of thermal inflation. We have also shown that, for the correct choice of model parameters, one can easily satisfy the Swampland criterion. Besides, for currect choice of equation of states($w_{re}$), one can also satisfy the recently proposed Trans-Planckian Censorship Conjecture(TCC).
\end{abstract}
\pacs{98.80.Cq}
\maketitle

\vspace{0.001in}


\baselineskip=15.4pt

\vspace{1cm}



\section{Introduction}\label{sec:intro}
\setcounter{equation}{0}
\setcounter{figure}{0}
\setcounter{table}{0}
\setcounter{footnote}{1}
 Cosmological inflation is a paradigm added to the standard hot big bang model, justifying a rapid exponential expansion of the Universe which is required to solve some of the initial condition problems such as `horizon' and `flatness' problem. Apart from the initial motivation to add an inflationary era, it was realised that it is essential to justify the structure formation of the Universe. After the first proposed model of inflation by Guth\cite{Guth}  several different models have been proposed\cite{,Linde,Liddle1,Steinhardt,Albrecht}. However, with the recent development of observational cosmology, it has became very hard for many of the models to survive the hard reality of observations. It has been argued in some places that inflation is a theoretical paradigm that cannot be refuted by any means which has been advocated by people like Steinhardt\cite{Ijjas}. But with the recent theoretical constraints coming from the String theorists, going by the name of {\bf S}wampland {\bf C}onjectures (SC) \cite{ArkaniHamed, Obied,Agrawal,Kinney} or the {\bf T}rans- {\bf P}lanckian {\bf C}onjecture(TCC) \cite{Bedroya,Bedroya1}, the whole idea of inflationary cosmology is in an existential crisis. One can obviously argue against the whole notion on which these conjectures are proposed. But String theory being our best hope to have a complete theory including gravity, one might need to be more judicious to even refute or bypass those claims. In this work, we refrain ourselves from making any comment on the validity of these conjectures. Rather, with a phenomenologically motivated class of inflationary model, we try to understand the ways to satisfy all the theoretical and observational demands, staying in the premise of the single field models. Single field models of inflation is at loggerheads with both the conjectures at least in the standard cold inflationary scenario. However, in some non-standard scenerios like warm inflation, we can also evade the SC. For detailed study readers are suggested to go through \cite{Motaharfar,Kamali,Brandenberger11}.  On the other hand, the observational data\cite{Hinshaw,Ade,Aghanim,Akrami} is pointing towards a plateau like potential for the inflationary potential. On that note, Dimopoulos and Owen, made an attempt to propose a phenomenological model dubbed as the power law plateau model in the standard cold inflationary scenario. However, it is realised in their paper that to have a successful observationally viable model, there is a need of effective number of e-folds ($N_e$) as low as $N_e= 35$. To satisfy that, they have introduced a phase of thermal inflation for lower $N_e$ to be allowed. In this paper, we tried to appreciate their model in the Randal-Sundrum brane-world\cite{Randall} scenario due to the reasons listed below:
 \begin{itemize}
 
 \item Power law plateau model realised in the RS $II$ brane world\cite{Randall1}, is much more robust in the sense that to satisfy the observational bounds on the inflationary observable, there is no need to put  by hand the phase of thermal inflation.
 \item It can give enough number of e-folds to solve the initial motivation of inflation: the flatness problem.
 \item For almost the whole class of this model in the braneworld, the SC can be well evaded for the correct choice of parameters keeping the observational bounds in mind.
 \item For particular choices from these class of models, even the TCC can be satisfied.
 
 \end{itemize}
 \subsection{RS $II$  Braneworld}
 The thermal history of the Universe can differ from the standard scenario once one considers the braneworld scenario\cite{Binetruy00,shiromizu,ida1,Fichet}. The original motivation of such scenario is to solve the hierarchy problem. One of such braneworld model is Randall-Sundrum model. The RS $II$ model is considered in our analysis. One of the immediate effects of considering this scenario is the modification of the Friedmann equation. In three brane, the cosmological expansion can be formulated by a more generalized Friedmann equation for an observer which is described by:
\begin{equation}
\left(\frac{\dot{a}}{a}\right)^2
=\frac{8 \pi G_{\rm N}}{3} \rho
-\frac{K}{a^2}+\frac{\Lambda_{4}}{3}
+\frac{\kappa_{5}^4}{36}\rho^2 + \frac{\mu}{a^4}~~.
\label{Friedmann}
\end{equation}
 where $a(t)$ is scale factor at time $t$ and $\rho$ is matter density in $3$ Dimensional (3-$D$) space, $G_{N}$ is the $4$ Dimensional  (4-$D$) Gravitational constant and it is related to $\kappa_5$ which is $5$ Dimensional (5-$D$) Gravitational constant. The last two terms are absent in the standard Friedmann equation. The last term plays key role in the radiation dominated era and is dubbed as the `dark radiation'. For details reader is advised to study references \cite{Ichiki, Sasankan, Sasankan1}.
 The fourth term plays key role in the inflationary era and would be the prime focus of our analysis. Again for details reader is advised to study the references \cite{Langlois, Gangopadhyay, Okada, Okada1, Calcagni, Calcagni1,Sabir,Sabir1}. 
 In the high energy limit when the Universe is dominated by the scalar field, one can rewrite the Friedmann equation as:
 
\begin{equation}
H^2 = \frac{V(\phi)}{3M_{P}^{2}} \biggl(1+\frac{V(\phi)}{\rho_0}\biggr)~~.
\label{hubble}
\end{equation}

Here $V(\phi)$ is the potential and $\phi$ is the field and $\rho_0$ is the variable which directly depends upon the $M_5$ ( $5$ Dimensional  Planck Mass) and $\rho_0$ is expressed as:

\begin{equation}
\rho_0 = 12 \frac{M_{5}^{6}}{M_{P}^{2}}~~.
\label{rho}
\end{equation}
The usual slow roll conditions are modified due to the modification in the evolution equation and this will correspond to the standard cosmology for $V/ \rho_{0}<<1$. In the subsequent sections, we will calculate everything in the limit  $V/ \rho_{0}>>1$. As the inflation is only restricted to the brane, the scalar perturbation behaves in similar fashion as the evolution of the standard cosmology after considering the altered Hubble expansion. The presence of extra dimension reshapes the tensor spectrum and gravitational power spectrum \cite{Tsujikawa1,Andriot}. The altered slow roll parameter can be written as:

\begin{equation}
\epsilon_{RS} =\frac{{ln(H^2)}'{V}'}{6H^2}~~,~~~\eta_{RS} = \frac{{V}''}{3H^2}~~.
\end{equation}
Here, the prime stands for the derivative with respect to $\phi$. Thus the observables such as $P_s$ (Scalar Perturbation), $n_s$ (Spectral  Index) and $r$ (Tensor to Scalar Ratio) gets modified and the slow roll analysis also gets modified accordingly, $P_s$ and $n_s$ can be denoted as: 

\begin{equation}
P_s = \frac{9}{4\pi^2}\frac{H^6}{V'^2}~~,~~~~~n_s = 1 - 6 \epsilon_{RS} + 2 \eta_{RS} ~~.~~
\label{Ps}
\end{equation} 

In a similar way, tensor power spectra changes as:

\begin{equation}
P_T = 8\biggl(\frac{H}{2\pi}\biggr)^2F(x_0)^2~~,
\label{Pt}
\end{equation}

where the extra factor $F(x_0)$ corresponds to 

\begin{equation}
F(x) = \biggl(\sqrt{1+x^2}  - x^2 \ln{\biggl[\frac{1}{x} + \sqrt{1+\frac{1}{x^2}}\biggr]}\biggr)^{-1/2}~~,
\label{fx}
\end{equation}
with $x=x_0 = 2(3H^2/\rho_0)^{1/2} $. This will reduce to standard cosmology for  $x_0\ll 1$. For $x_0\gg1$, the extra factor can be approximated by  $\sqrt{3x_0/2}$. The tensor to scalar ratio can be written as:

\begin{equation}
  r = P_T/P_s 
\label{r}
\end{equation}


\subsection{Swampland Conjecture $\&$ TCC}
Swampland Criteria is proposed by Vafa et al,  \cite{Vafa:2005} to
evade any (meta-)stable de Sitter constructions within String landscapes. These conjectures make the paradigms of accelerations such as the inflationary epoch or the quintessence dark energy, extremely difficult to survive. It has been shown that if the  two conjectures are true, that can rule out all the single field inflationary models at least in the standard cosmology. In recent times, another conjecture is proposed, namely the {\bf T}rans- {\bf P}lanckian {\bf C}onjecture (TCC) \cite{Brahma1, Martin1, Brandenberger1} which puts these paradigms of single scalar field dominated accelerated expansion theories in more trouble. On that note, the two SC conjectures are \cite{Palti, Brahma, Brahma2}:
\begin{itemize} 
   \item {\it SC1:} The range traversed by scalar fields in field space has the maximum limit:
   \begin{equation}
    \bigg|\frac{\Delta \varphi}{M_{Pl}}\bigg|\leqslant \Delta \sim \mathcal{O}(1)~.
    \end{equation}
 
   \item {\it SC2:} This criterion limits the gradient of scalar potentials in an effective field theory as:
    \begin{equation}
        M_{Pl} \frac{|V'|}{V} \geqslant c \sim \mathcal{O}(1)
   \end{equation}
\end{itemize} 
{\it SC2}  is in direct conflict with the idea of slow roll inflation\cite{Garg,Ooguri}. As we know, the slow roll inflation requires 
the first slow roll parameter $\epsilon_{s}$, defined as: $\epsilon_{s} = \frac{M_{Pl}^2}{2}(V'/V)^2$,  to be less than $1$, to have a successful inflationary epoch. It is quite evident that the {\it SC2} and the slow roll requirements are at loggerheads. 

Finally, the TCC demands following \cite{Bedroya} ``Any inflationary model which can stretch the quantum  fluctuations with wavelengths smaller than the Planck scale out of the Hubble horizon is in the swampland''. This can be realised in terms of the number of e-folds ($N$) as:
\begin{equation}
Exp[N] < \frac{M_{Pl}}{H_e}~,
\end{equation}
where $H_e$ is the Hubble expansion rate at the end of inflation. In \cite{Mizuno}, it is  shown in a model independent way that there is an universal upper bound on the inflationary Hubble expansion rate ($H_{inf}$) fixed by the reheating temperature $T_{re}$ as:
\begin{equation}
\frac{H_{inf}}{M_{Pl}}\leqslant \frac{T_0}{T_{re}}~,
\end{equation}
where, $T_0$ is the CMB temperature today. 

The {\it SC1} can be taken care of by keeping the field value of the inflaton to be  always sub-Planckian and in our analysis we have considered such field values.

 To satisfy the {\it SC2}, in the rest of the paper, we have rewritten $\epsilon_{RS}$ in terms of $\epsilon_{s}$  and satisfying conjecture {\it SC2}, implies $\epsilon_{s}  \sim c^2/2 \sim \mathcal{O}(1)$. However, the redefined  $\epsilon_{RS}$ can be less than one to successfully carry out the inflation. The rewritten form is given as:
\begin{equation}
\epsilon_{RS} =\frac{{ln(H^2)}'(2\epsilon_{s})^{1/2}V}{6H^2}~~
\label{eqsc2}
\end{equation}

For TCC, following \cite{Mizuno,Tenkanen}, we can consider the conclusion of their paper for the equation of state during reheating ($w_{re}=-1/3$) and write:
\begin{equation}
r\leqslant 2 \times 10^{-8}\bigg(\frac{1 MeV}{T_{re}}\bigg)^2
\label{mmm}
\end{equation}


\subsection{Power-Law Plateau Type Potential}

\label{sec:sect}
In Supersymmetry, considering $\phi_1$,  $\phi_2$ and $S$ as the chiral superfileds as done in \cite{Dimopoulos}, the superpotential can be written as:
\begin{equation}
W= \frac{S(\phi_1^2-\phi_2^2)}{2m}~,
\end{equation}
where $m$ is the sub-Planckian scale. Finally, one can get the power law plateau (PLP) kind potential as:
\begin{equation}
V= \frac{M^4 \phi^2}{m^2+ \phi^2}~,
\label{n2q2}
\end{equation}
where $M$ is the GUT scale. Thus, one can define a class of model which has the same feature as that in (\ref{n2q2}). The class of potentials can be written as:
\begin{equation}
V = V_0\biggl(\frac{{\phi}^n}{{\phi}^n+{{m}^n}}\biggr)^q
\label{Pot}
\end{equation}
Here $m$ is the mass scale, $\phi$ is the real scalar field. Generally,  $n$ and  $q$ are the real parameters,  $V_{0}$ is the scale of inflation. Here we assume $\phi > m$ to maintain the shape of the plateau potential, otherwise it is reduced  to  the monomial  inflation model \cite{Lin} with  $V \propto \phi^{nq}$. In the following discusion, we take the assumption $\frac{m}{\phi}\ll 1$. We vary $n$, $q$  from 1 to 4. This gives us 16 different combination for the different values of  $n$, $q$ but here we restrict our discussion to only those combinations which satisfy both the Swampland conjuctures. Furthermore, we  take $\epsilon_{s} = 1$ which is the second swampland conjecture (SC2). In our analysis the initial value  $\phi_i$ of the field is always less than $M_{Pl}$, thus it is obvious that the (SC1) is maintained throughout the analysis. 
 In  section II, a detailed analysis of PLP inflationary potential in the RS $II$ is carried out. In  section III, the reheating phase after the inflation is discussed. Then in section IV, we have discussed the TCC in the context of our analysis.  Finally we have drawn our conclusion in the last section V.


\section{Analysis of inflationary observables}
\setcounter{equation}{0}
\setcounter{figure}{0}
\setcounter{table}{0}
\setcounter{footnote}{1}

First in this section first we have calculated the inflationary observables in terms of the model parameter for the PLP in the braneworld scenario.

\begin{figure}
    \centering
    \subfigure[]{\includegraphics[width=0.4\textwidth]{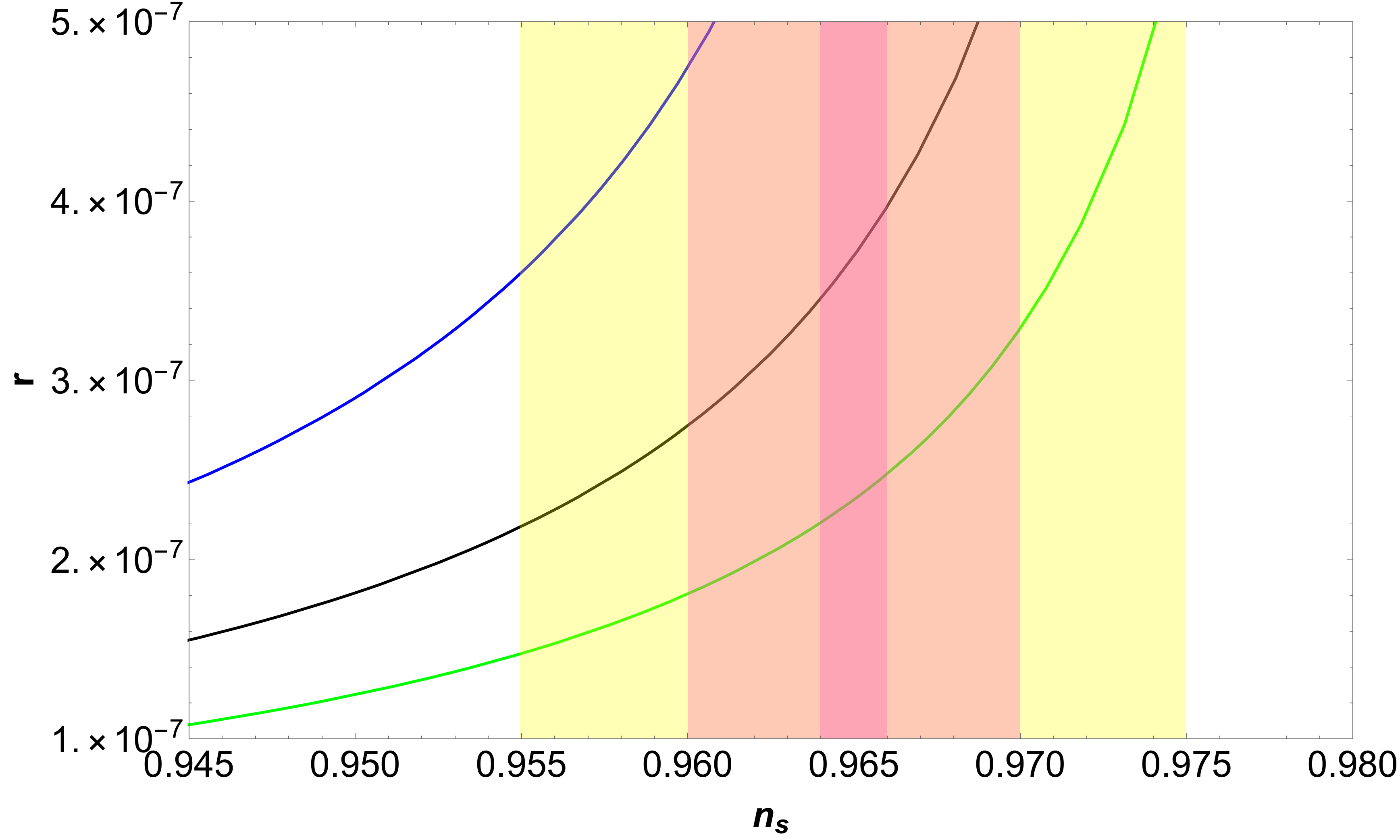}} 
    \subfigure[]{\includegraphics[width=0.4\textwidth]{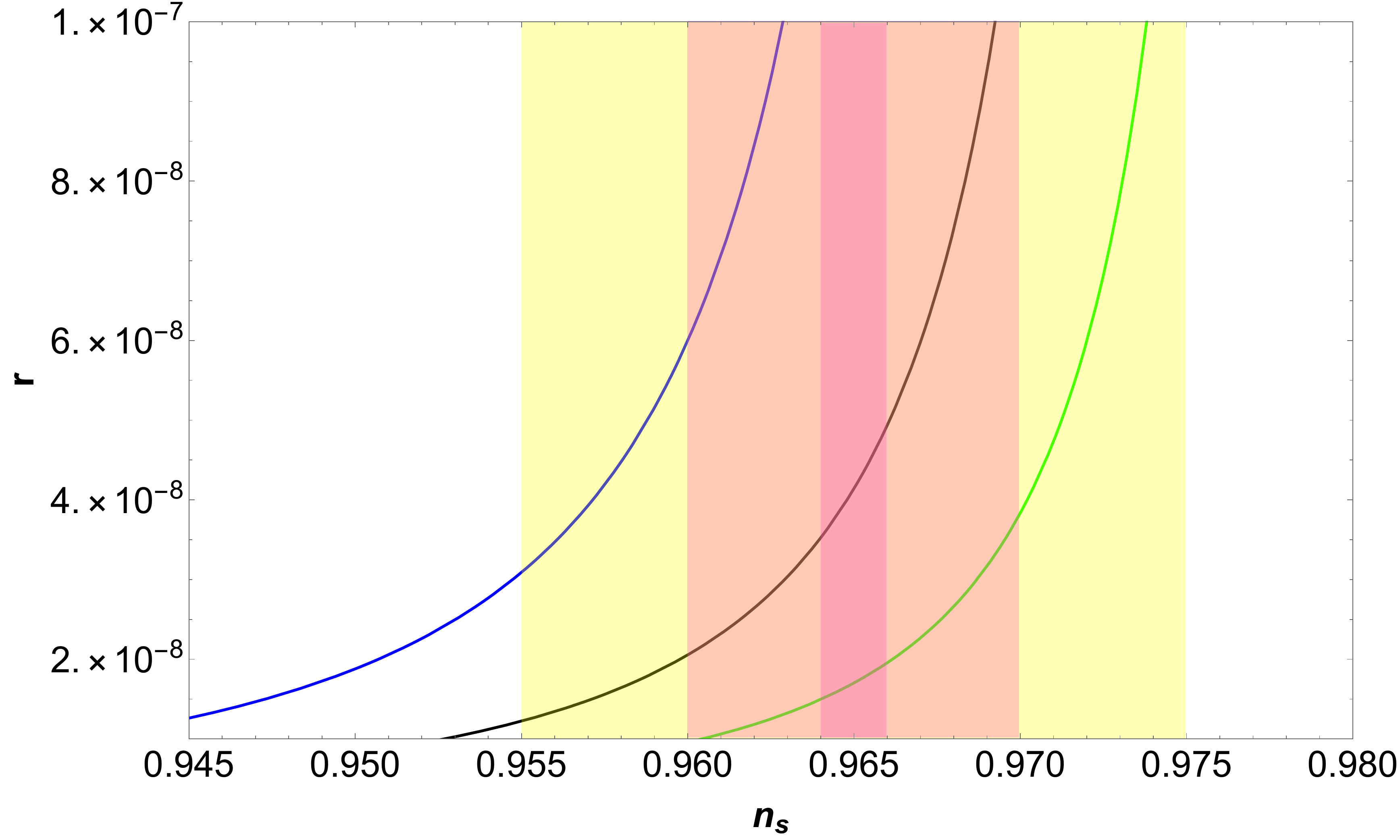}} 
    \subfigure[]{\includegraphics[width=0.4\textwidth]{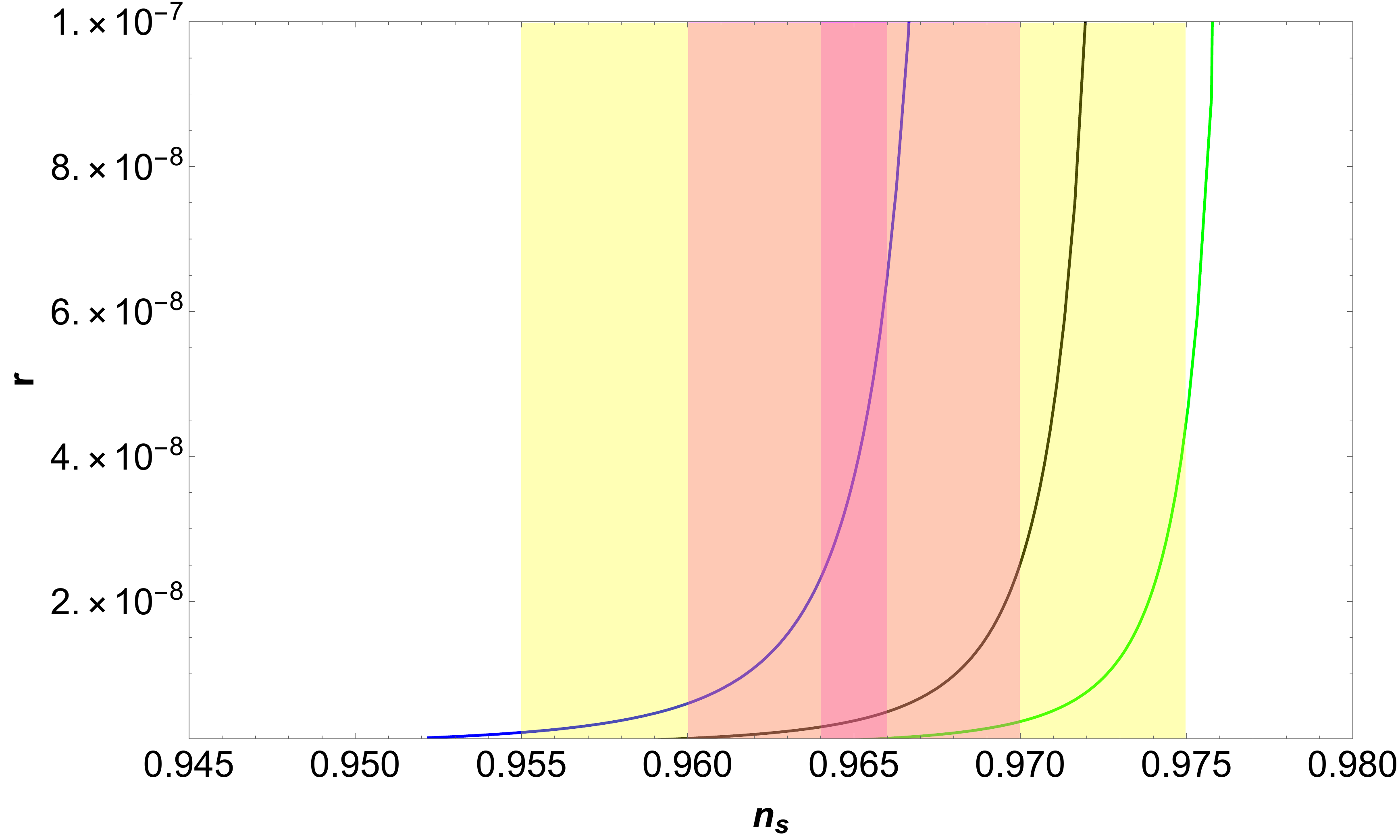}}
    \subfigure[]{\includegraphics[width=0.4\textwidth]{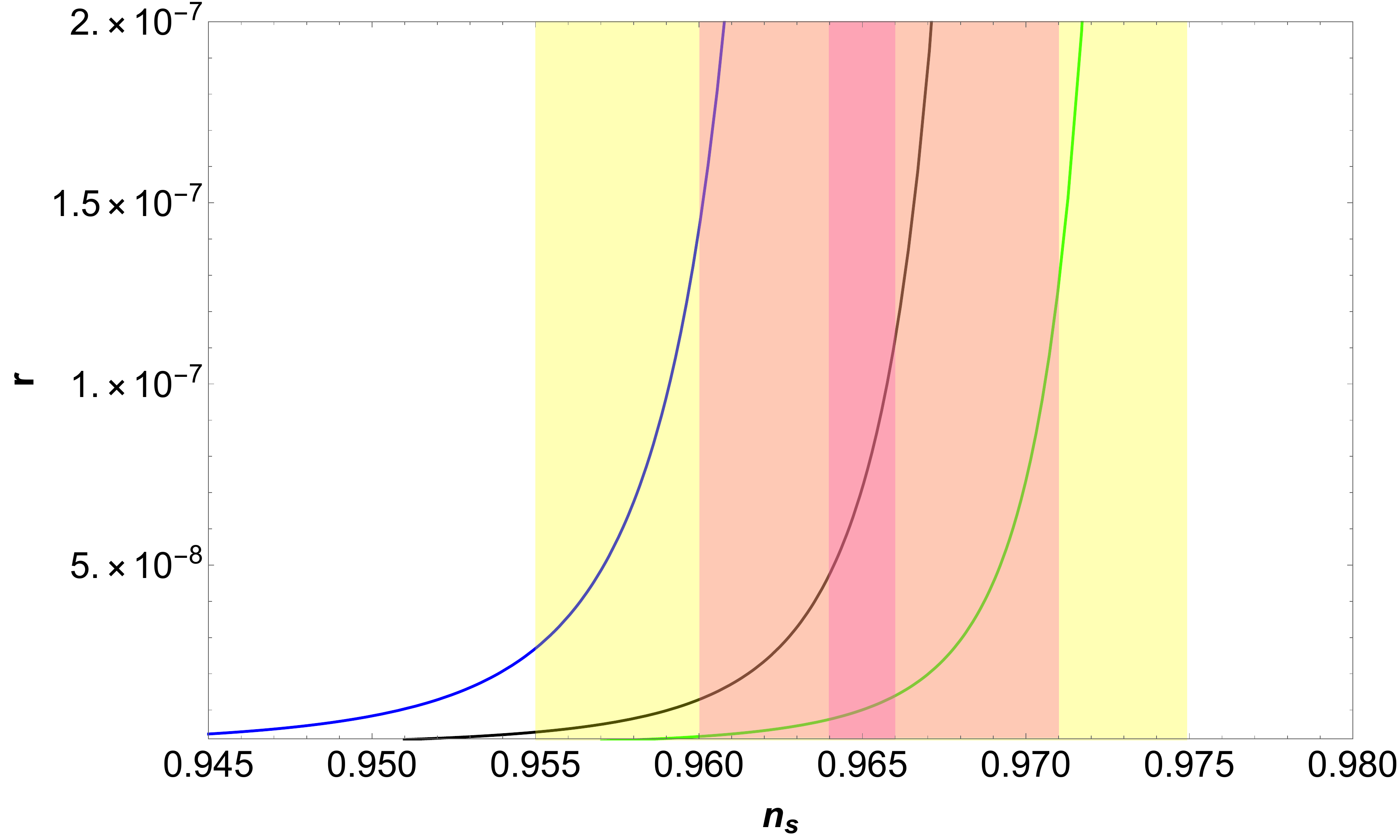}}
    \caption{Plots of $r$ and $n_{s}$ as a function $\rho_{0}$ for fixed values of $V_0 = 10^{8}\rm GeV $. In fig (a). $n=1,q=1$ and $m=10^{-6}$, fig (b). $n=2,q=1$ and $m=10^{-4}$,  fig (c). $n=3,q=1$ and $m=10^{-3}$ and  in fig (d). $n=4,q=1$ and $m=10^{-3}$. The blue line corresponds to $N_{e} = 55$, the black line corresponds to $N_{e} = 65$, the green line corresponds to $N_{e} = 75$ . The light pink shaded region corresponds to the 1-$\sigma$ bound and the yellow shaded region is 2-$\sigma$ bound on $n_s$ from {\it Planck'18} [TT,TE, EE+lowE+lensing+BK15].  The dark pink shaded region corresponds to the 1-$\sigma$ bound of future CMB observations  using same central value for $n_s$\cite{prism,euclid}.  }
    \label{figq1}
\end{figure}

\begin{figure}
    \centering
    \subfigure[]{\includegraphics[width=0.4\textwidth]{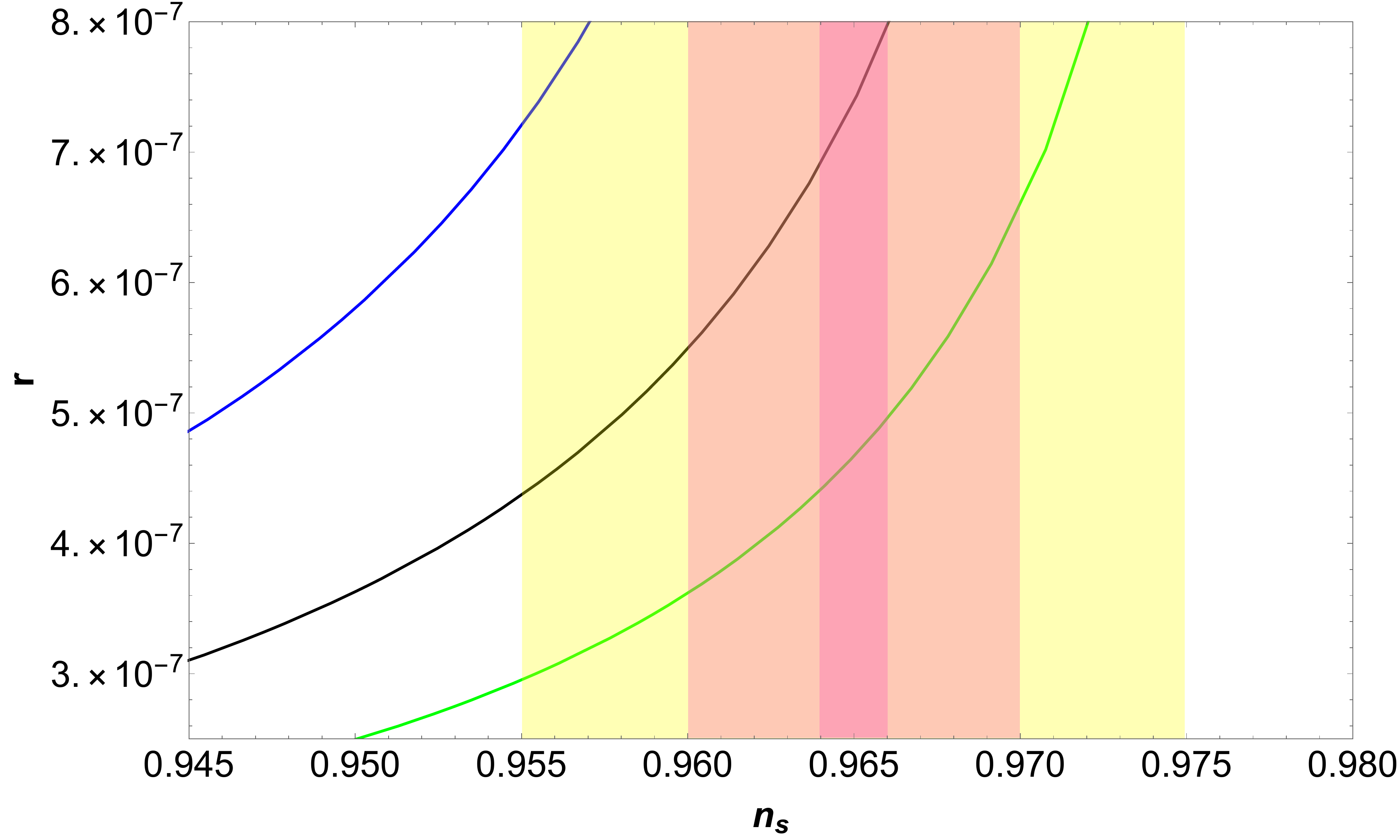}} 
    \subfigure[]{\includegraphics[width=0.4\textwidth]{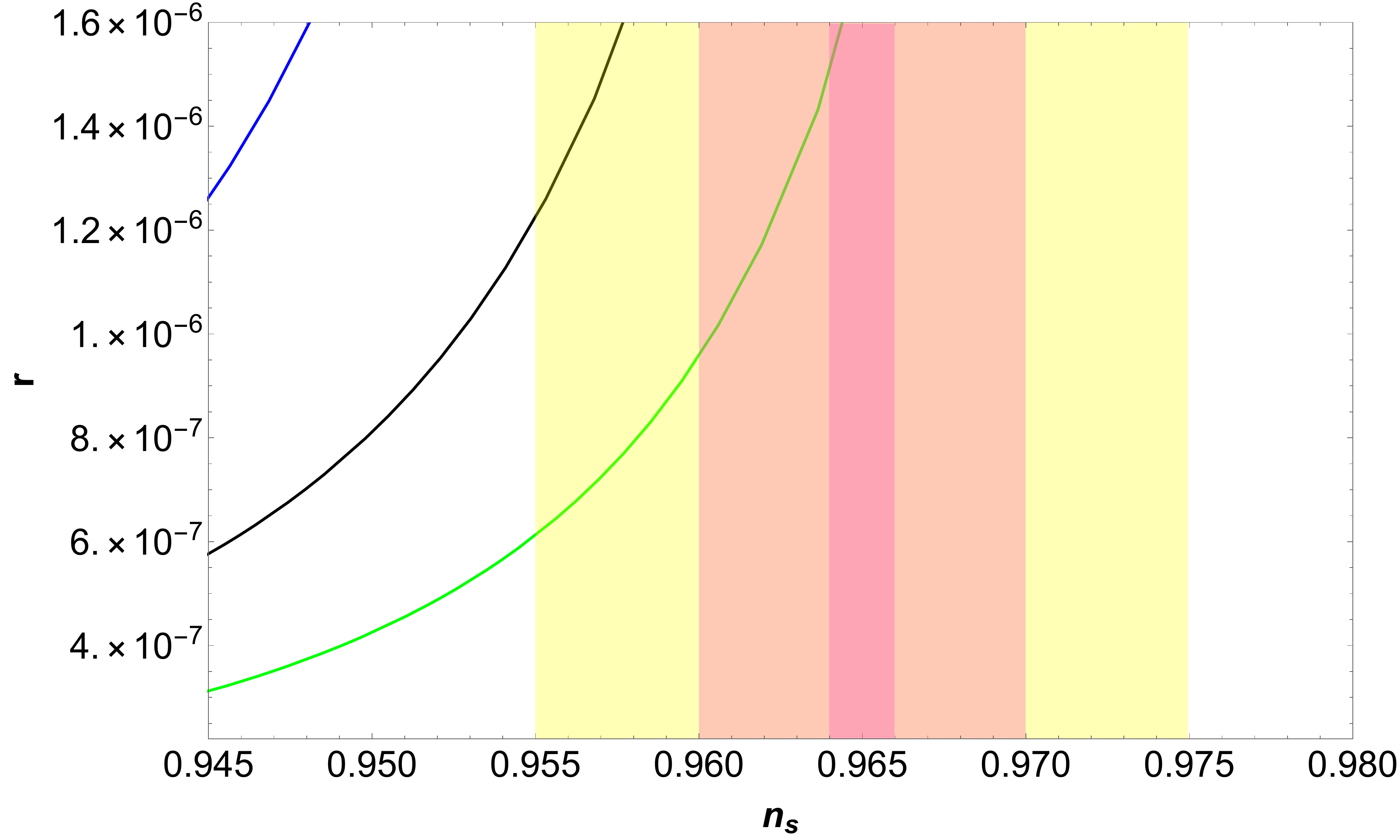}} 
    \subfigure[]{\includegraphics[width=0.4\textwidth]{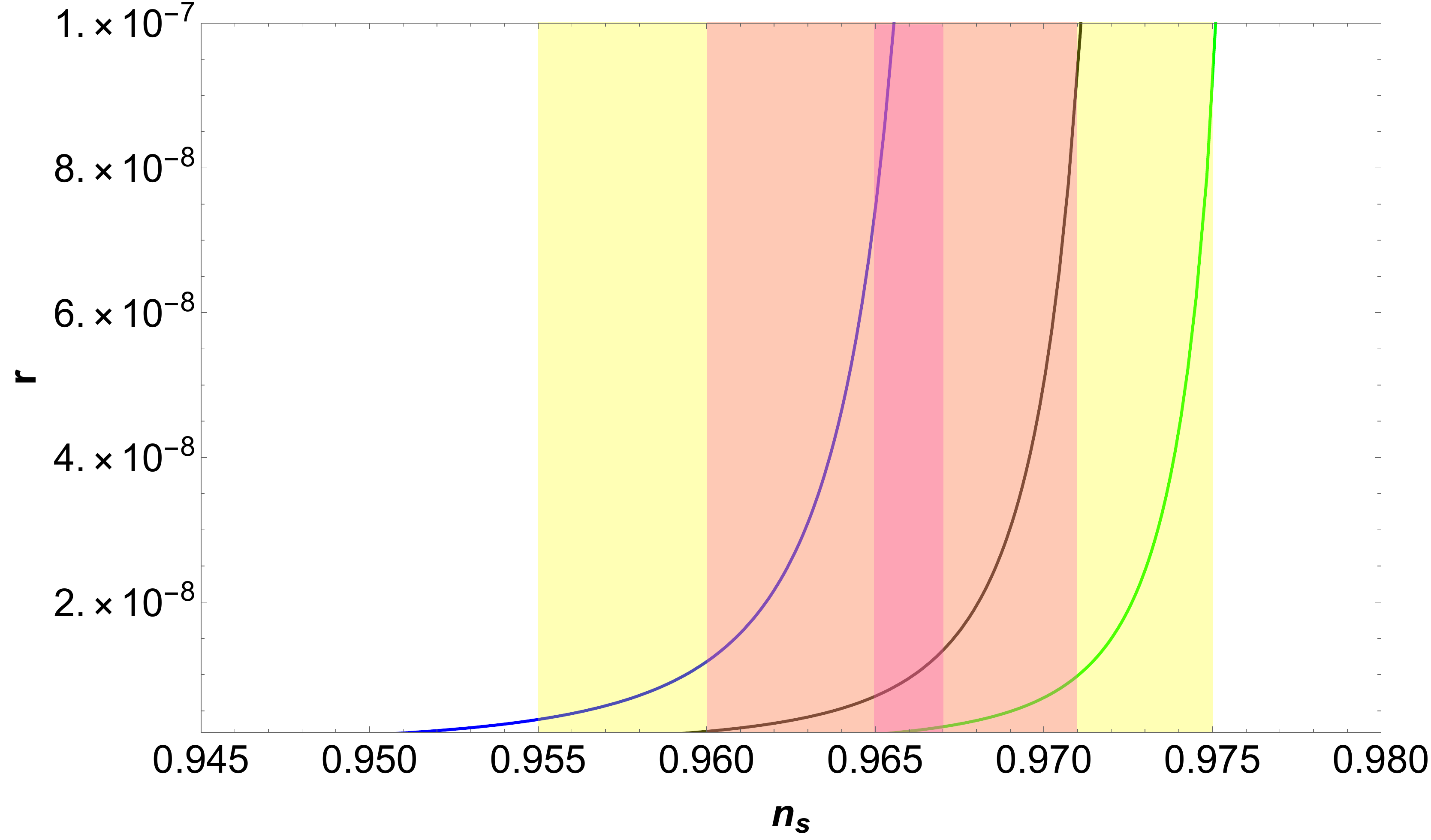}}
    \subfigure[]{\includegraphics[width=0.4\textwidth]{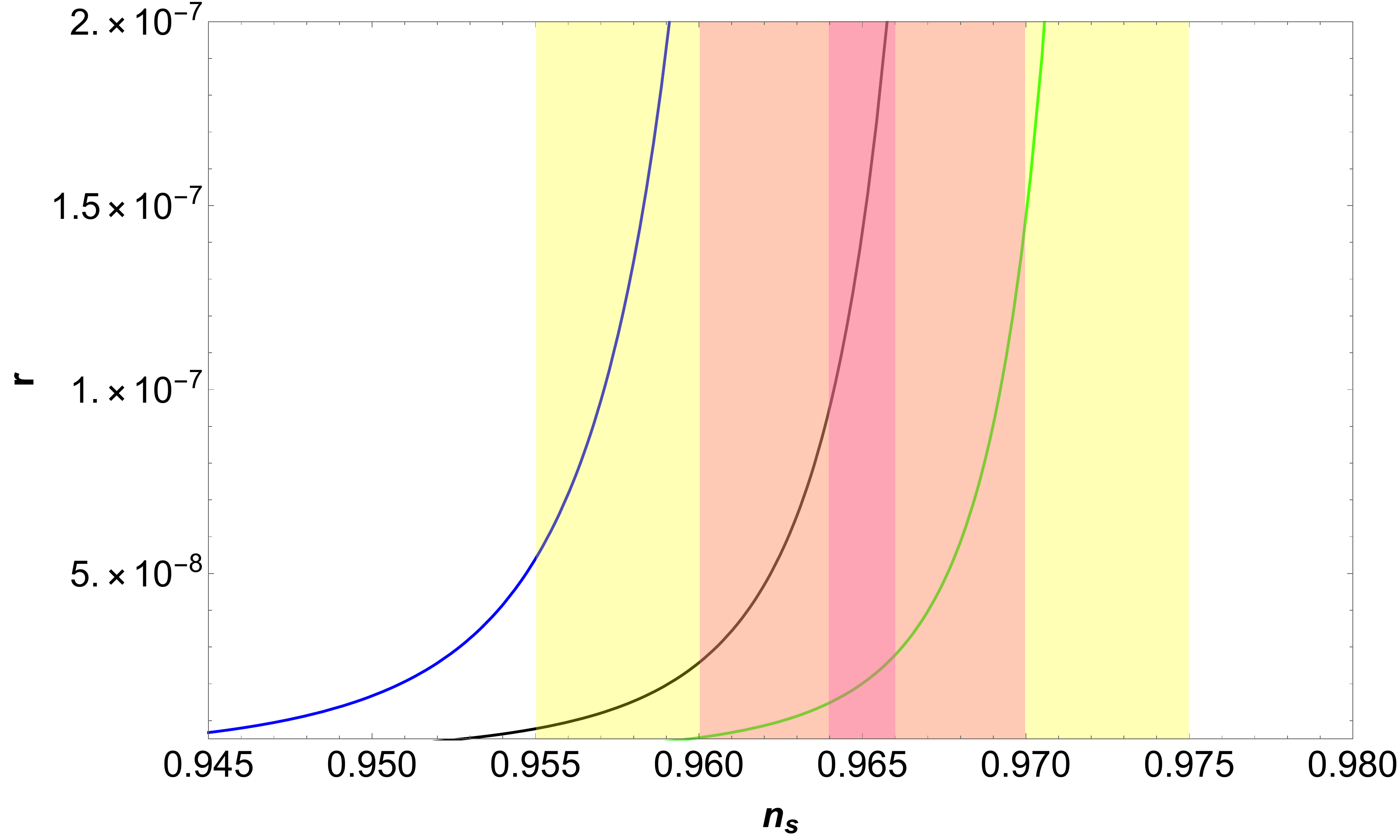}}
    \caption{Plots of $r$ and $n_{s}$ as a function $\rho_{0}$ for fixed values of $V_0 = 10^{8}\rm GeV $. In fig (a). $n=1,q=2$ and $m=10^{-6}$, fig (b). $n=2,q=2$ and $m=10^{-3}$,  fig (c). $n=3,q=2$ and $m=10^{-3}$ and  in fig (d). $n=4,q=2$ and $m=10^{-3}$. The blue line corresponds to $N_{e} = 55$, the black line corresponds to $N_{e} = 65$, the green line corresponds to $N_{e} = 75$ . The light pink shaded region corresponds to the 1-$\sigma$ bound and the yellow shaded region is 2-$\sigma$ bound on $n_s$ from {\it Planck'18} [TT,TE, EE+lowE+lensing+BK15].  The dark pink shaded region corresponds to the 1-$\sigma$ bound of future CMB observations  using same central value for $n_s$\cite{prism,euclid}.  }
    \label{figq2}
\end{figure}

\begin{figure}
    \centering
    \subfigure[]{\includegraphics[width=0.4\textwidth]{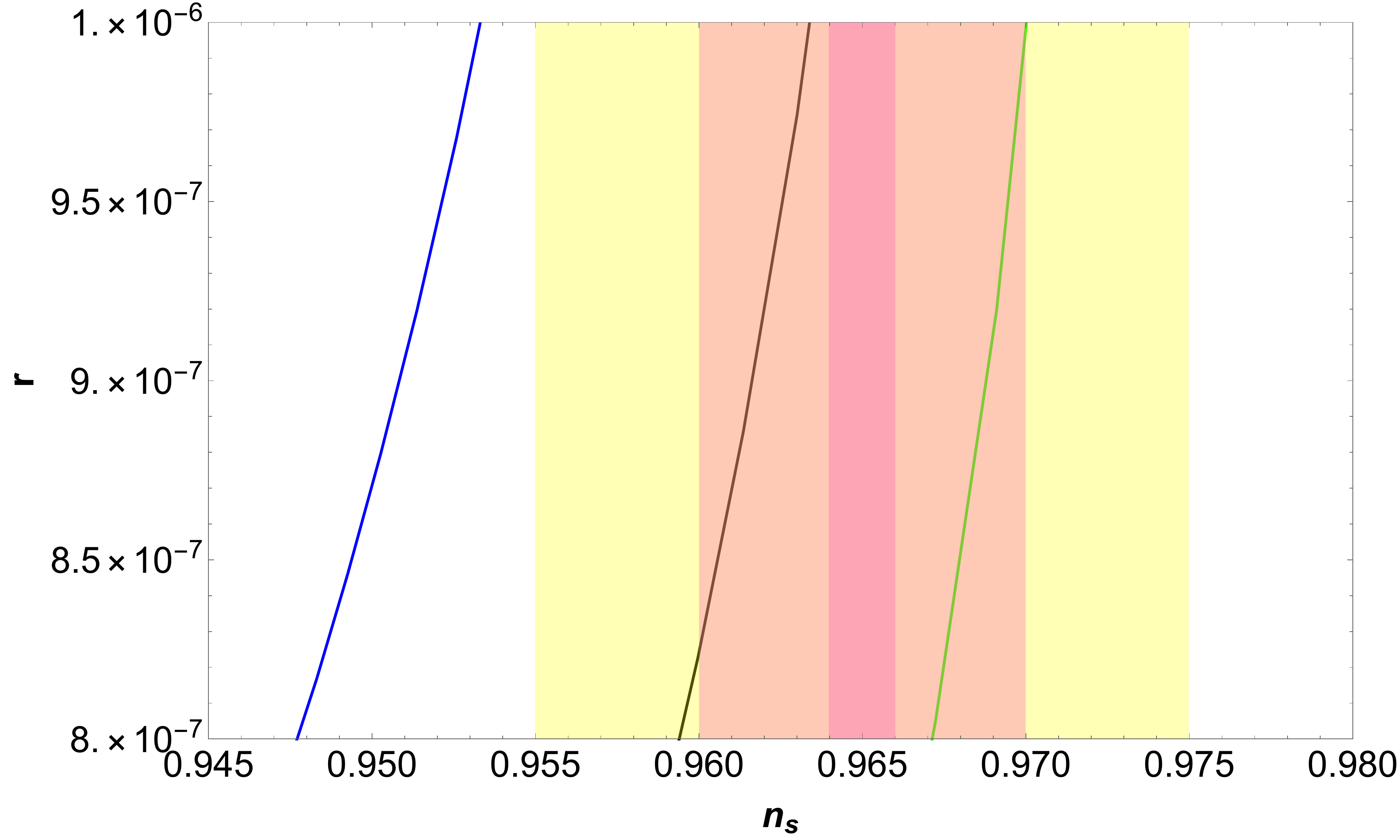}} 
    \subfigure[]{\includegraphics[width=0.4\textwidth]{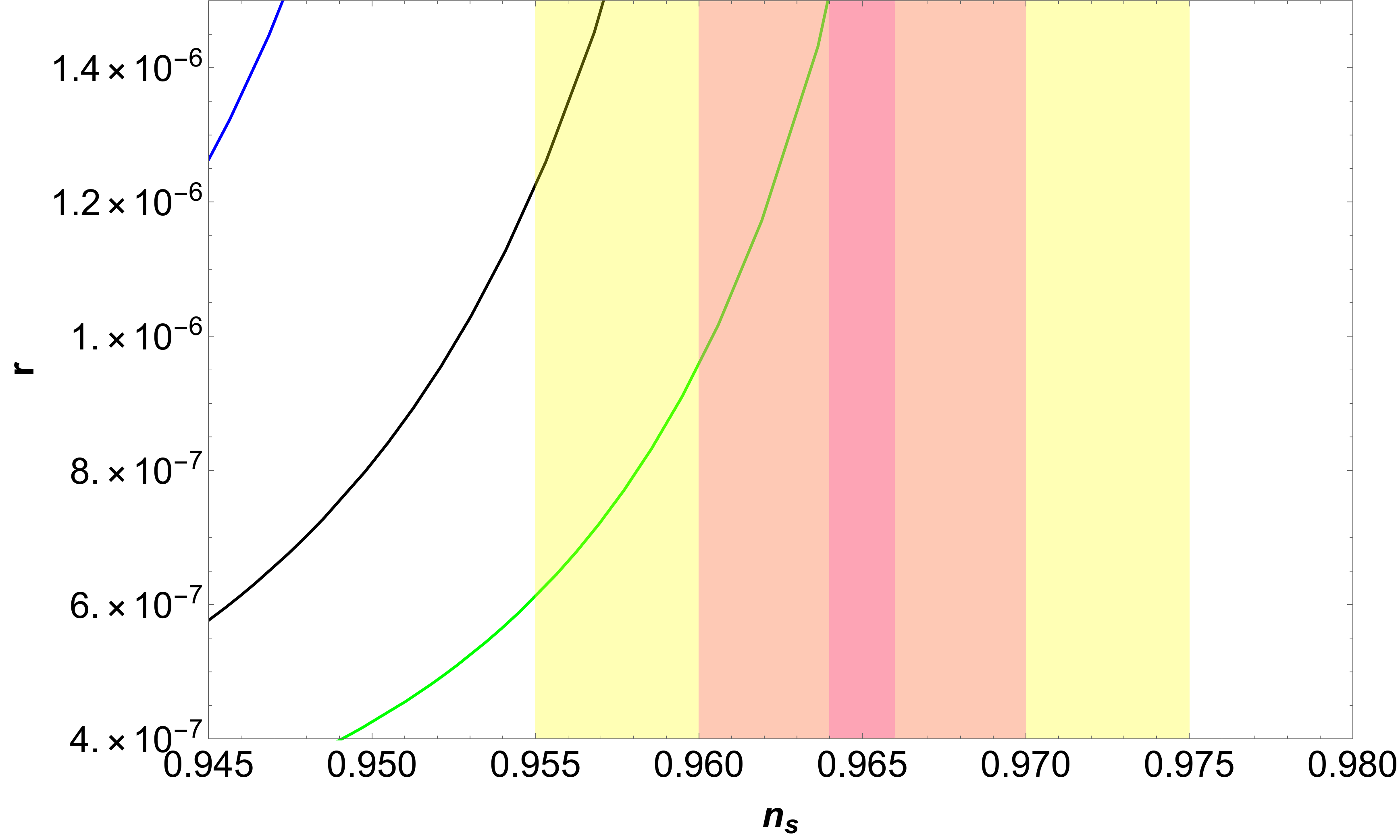}} 
    \subfigure[]{\includegraphics[width=0.4\textwidth]{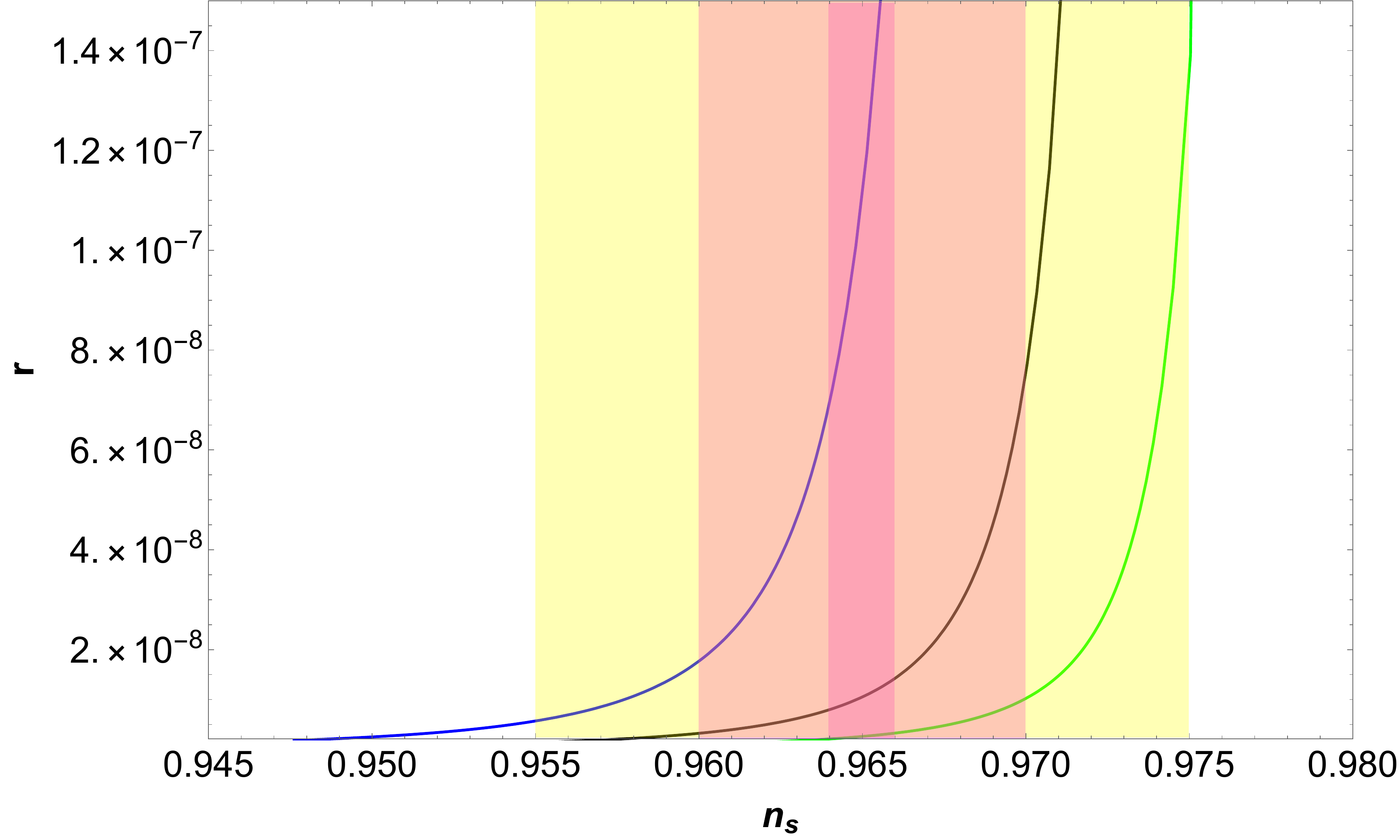}}
    \subfigure[]{\includegraphics[width=0.4\textwidth]{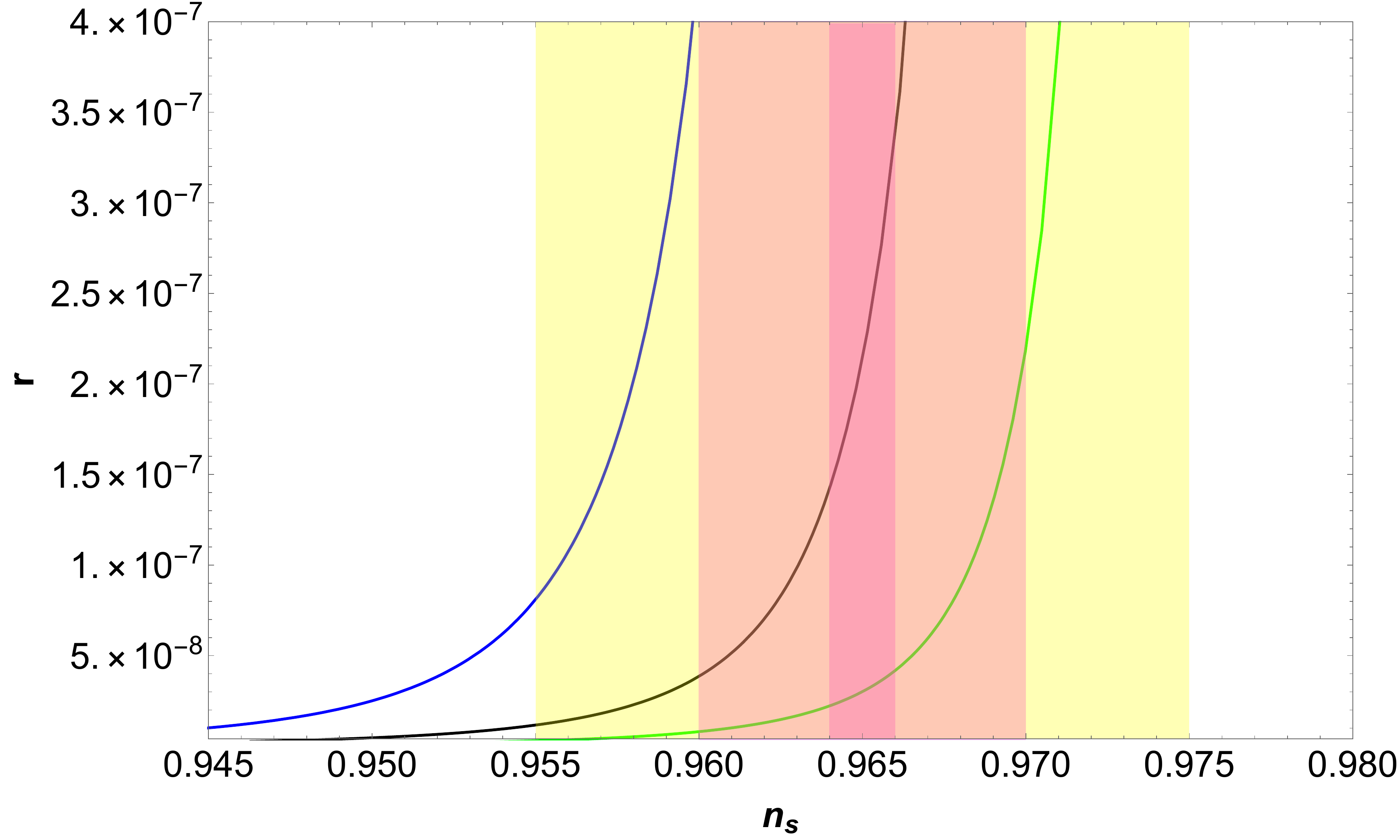}}
    \caption{Plots of $r$ and $n_{s}$ as a function $\rho_{0}$ for fixed values of $V_0 = 10^{8}\rm GeV $. In fig (a). $n=1,q=3$ and $m=10^{-6}$, fig (b). $n=2,q=3$ and $m=10^{-3}$,  fig (c). $n=3,q=3$ and $m=10^{-3}$ and  in fig (d). $n=4,q=3$ and $m=10^{-3}$. The blue line corresponds to $N_{e} = 55$, the black line corresponds to $N_{e} = 65$, the green line corresponds to $N_{e} = 75$ . The light pink shaded region corresponds to the 1-$\sigma$ bound and the yellow shaded region is 2-$\sigma$ bound on $n_s$ from {\it Planck'18} [TT,TE, EE+lowE+lensing+BK15]. The dark pink shaded region corresponds to the 1-$\sigma$ bound of future CMB observations  using same central value for $n_s$\cite{prism,euclid} . }
    \label{figq3}
\end{figure}

\begin{figure}
    \centering
    \subfigure[]{\includegraphics[width=0.4\textwidth]{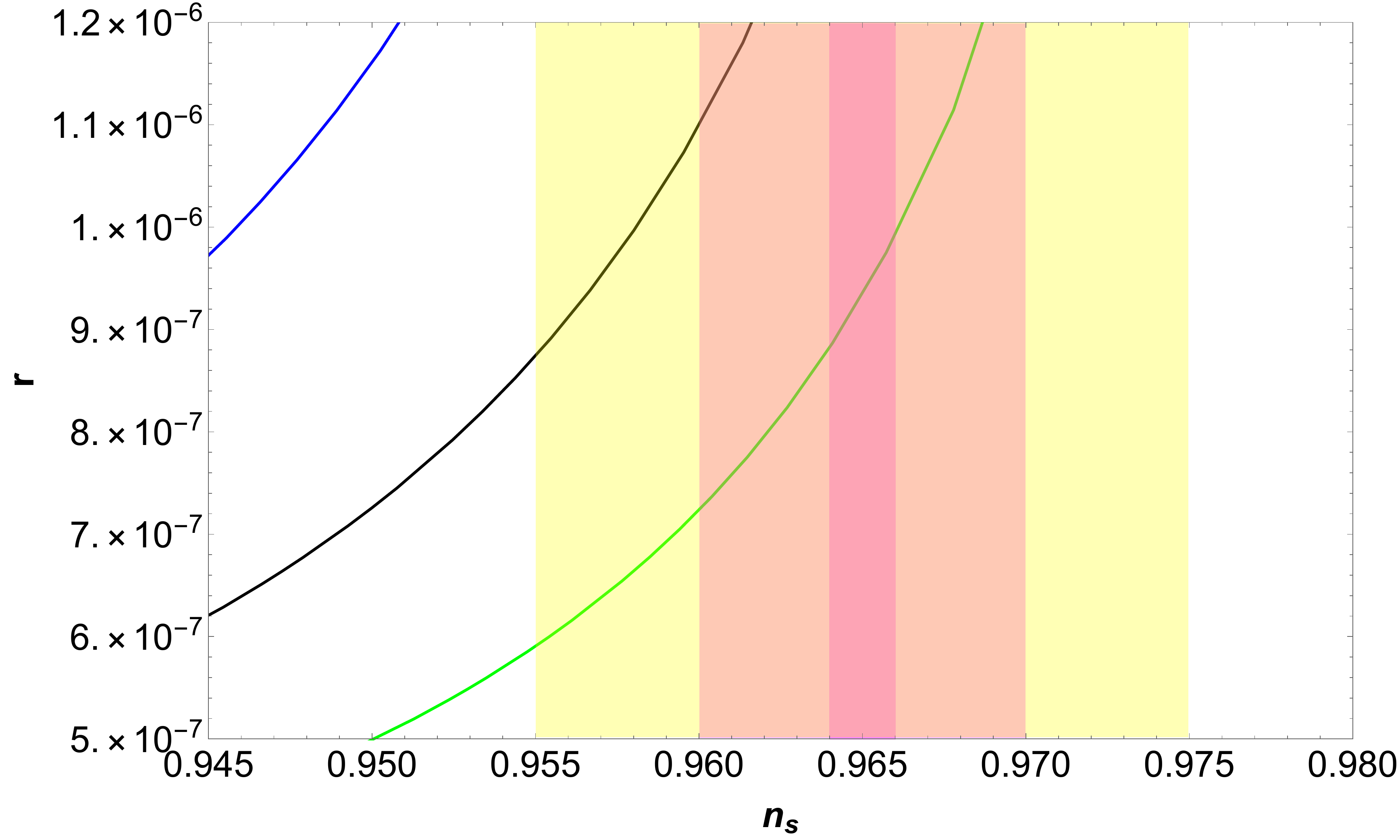}} \label{fig14}
    \subfigure[]{\includegraphics[width=0.4\textwidth]{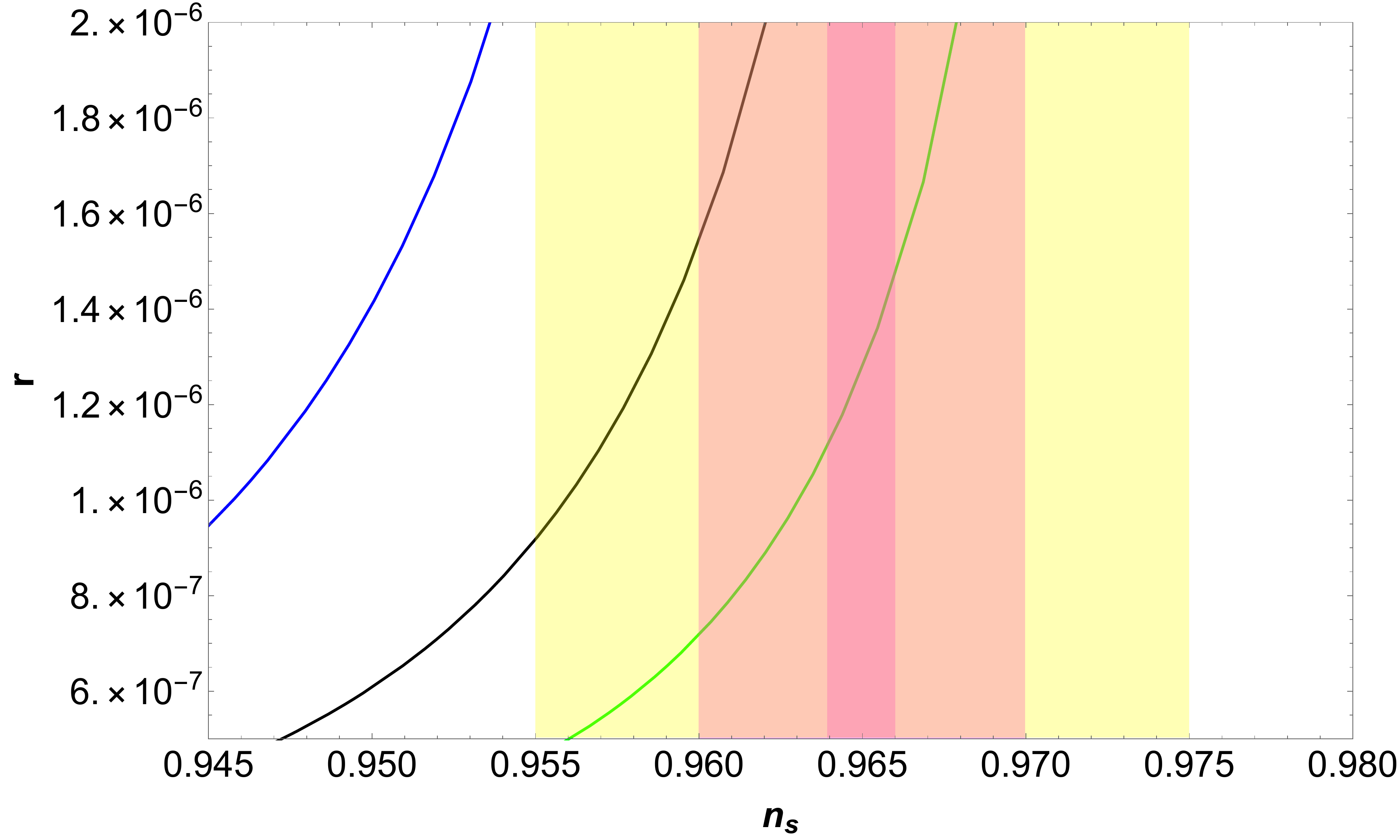}} \label{fig24}
    \subfigure[]{\includegraphics[width=0.4\textwidth]{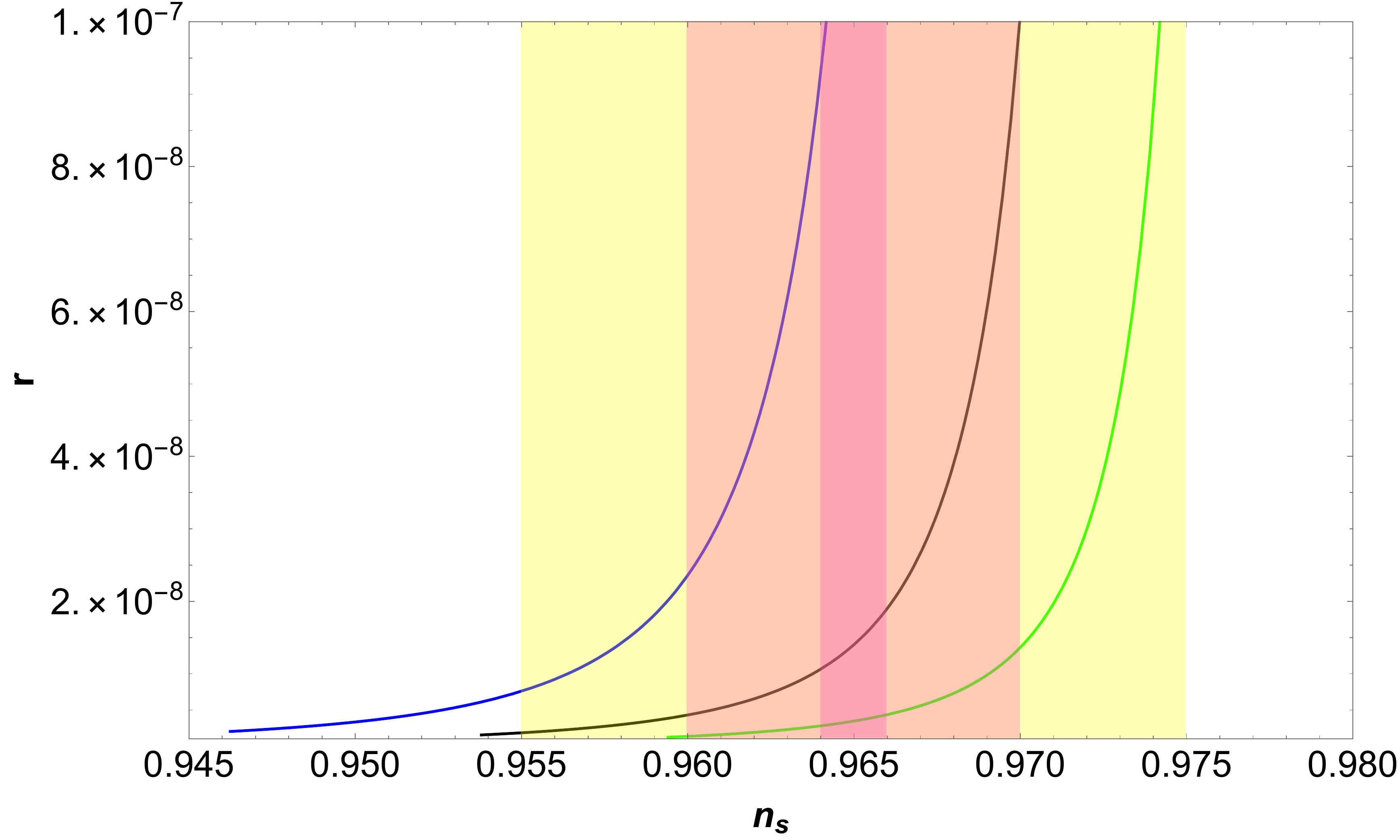}}\label{fig34}
    \subfigure[]{\includegraphics[width=0.4\textwidth]{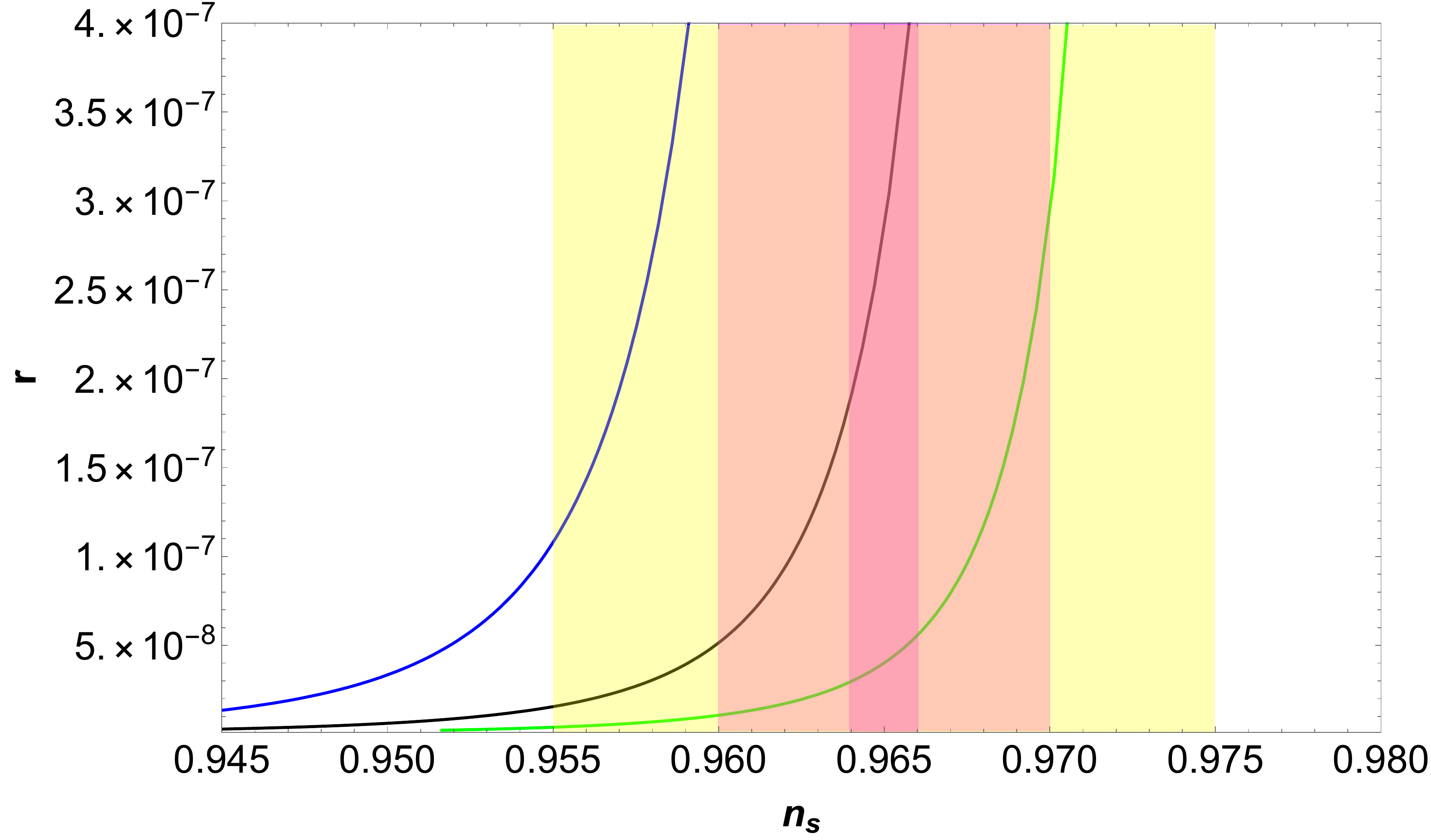}}\label{fig44}
    \caption{Plots of $r$ and $n_{s}$ as a function $\rho_{0}$ for fixed values of $V_0 = 10^{8}\rm GeV $. In fig (a). $n=1,q=4$ and $m=10^{-6}$, fig (b). $n=2,q=4$ and $m=10^{-3}$,  fig (c). $n=3,q=4$ and $m=10^{-3}$ and  in fig (d). $n=4,q=4$ and $m=10^{-3}$. The blue line corresponds to $N_{e} = 55$, the black line corresponds to $N_{e} = 65$, the green line corresponds to $N_{e} = 75$ . The light pink shaded region corresponds to the 1-$\sigma$ bound and the yellow shaded region is 2-$\sigma$ bound on $n_s$ from {\it Planck'18} [TT,TE, EE+lowE+lensing+BK15]. The dark pink shaded region corresponds to the 1-$\sigma$ bound of future CMB observations  using same central value for $n_s$\cite{prism,euclid} }
    \label{figq4}
\end{figure}

 Here, following Eq. (\ref{Ps}), (\ref{Pt}) and (\ref{r}) one can write the generalised formula for $n_s$ and $r$  as :

\begin{eqnarray}
n_s = 1-\frac{2 \; n\; q \;\rho_{0} \; m^n \left(\left(2^{\frac{1}{n+1}} \left(\frac{V_0  m^{-n}}{n q \rho_{0}  \sqrt{\epsilon_s}}\right)^{-\frac{2}{n+1}} \left(2^{\frac{1}{2 n+2}} \left(\frac{V_0  m^{-n}}{n q \rho_{0}  \sqrt{\epsilon_s}}\right)^{-\frac{1}{n+1}}\right)^n+\frac{n (n+2) {N} q \rho  m^n}{V_0 }\right)^{\frac{1}{n+2}}\right)^{-n-2}} {V_0 }\nonumber\\
\nonumber \\
 \times  \left\{ 3 \sqrt{2} \sqrt{\epsilon_s} \left(2^{\frac{1}{n+1}} \left(\frac{V_0  m^{-n}}{n q \rho_{0}  \sqrt{\epsilon_s}}\right)^{-\frac{2}{n+1}} \left(2^{\frac{1}{2 n+2}} \left(\frac{V_0  m^{-n}}{n q \rho_{0}  \sqrt{\epsilon_s}}\right)^{-\frac{1}{n+1}}\right)^n+\frac{n (n+2) {N} q \rho_{0}  m^n}{V_0 }\right)^{\frac{1}{n+2}}+n+1\right\}
\end{eqnarray}

\begin{eqnarray}
r = \frac{24 n^2 q^2 \rho_{0} \; m^{2 n} \left(\left(2^{\frac{1}{n+1}} \left(\frac{V_0  m^{-n}}{n q \rho_0  \sqrt{{\epsilon_s}}}\right)^{-\frac{2}{n+1}} \left(2^{\frac{1}{2 n+2}} \left(\frac{V_0  m^{-n}}{n q \rho_0  \sqrt{{\epsilon_s}}}\right)^{-\frac{1}{n+1}}\right)^n+\frac{n (n+2) {N} q \rho_0  m^n}{V_0 }\right)^{\frac{1}{n+2}}\right)^{-2 (n+1)}}{V_0 }
\end{eqnarray}
The interesting result to conclude from this section is the obvious effect of the bulk dimension which helps to tune the PLP model to get the inflationary parameters in the observationally allowed region without compromising with the satisfactory number of e-foldings. One can deduce that from fig.~(\ref{figq1}-\ref{figq4}). All the results quoted in Table~(\ref{Table}) satisfies the {\it SC2} through the previously stated condition by Eq.~(\ref{eqsc2}). {\it SC1} is satisfied from the fact that the initial condition is fixed such that the whole inflationary dynamics remains sub-Planckian.




\begin{table}
\begin{tabular}{|m{0.5cm}  | m{1cm}|   m{1.5cm}|   m{2.3cm}|  m{1cm}|   m{1.5cm}|  m{1.5cm}|   m{1.5cm}|}
\hline
$ q $    &$ n $   &$ m ({\rm GeV})$   &$M_{5}\times10^{16} {\rm GeV}$   &$ N $   &$ r \times10^{-6}$   &$ n_{s} $  &$ P_{s} \times10^{-9} $ \\ 
\hline
&     1&     $10^{12}$&    8.05&   65&    0.393&   0.965878&    $2.08$   \\ 

1&   2&    $10^{14}$&    9.50&   55&    0.049&   0.964624&    $2.08$   \\ 
 
&    3&    $10^{15}$&     10.0&   55&    0.024&   0.964117&     $2.08$    \\   
 
&     4&    $10^{16}$&    9.20&    65&   0.079&   0.965243&     $2.08$    \\
     \hline
&     1&    $10^{12}$&    7.84&   75&    0.543&   0.967437&     $2.07$    \\     
     
2&   2&   $10^{15}$&    7.45&   75&    0.927&   0.963407&     $2.22$    \\
     
&    3&    $10^{15}$&     10.0&   65&    0.024&   0.968557&     $2.06$    \\
     
&     4&     $10^{16}$&    8.93&   65&    0.112&   0.964426&     $2.09$    \\
     \hline
&     1&     $10^{12}$&   7.63&   75&    0.752&   0.966186&     $2.07$     \\     
     
3&    2&    $10^{15}$&   7.43&   75&    1.191&   0.962066&     $2.09$     \\
     
&     3&    $10^{15}$&     9.95&   65&    0.031&   0.968153&     $2.07$      \\
     
&     4&     $10^{16}$&   8.70&   65&    0.143&   0.964009&     $2.26$    \\
      \hline
 &     1&    $10^{12}$&   7.46&   75&    0.951&   0.965326&     $2.12$     \\         
      
4&     2&   $10^{15}$&   7.22&   75&    1.441&   0.965881&     $2.08$     \\    

&    3&    $10^{15}$&    9.83&   65&    0.036&   0.967838&     $2.04$    \\
     
&     4&   $10^{15}$&    8.68&   65&    0.158&   0.963523&     $2.09$    \\    
        \hline

\end{tabular}
\caption{Values of different cosmological observables for the different combination of $n$, $q$, $n$, $m$, $N$ and $M_5$}
\label{Table}
\end{table}



\section{Reheating Analysis}
\setcounter{equation}{0}
\setcounter{figure}{0}
\setcounter{table}{0}
\setcounter{footnote}{1}
\label{sec:reh}

After the end of the inflation an epoch of reheating is required to have the approximate temperature of the universe to start the BBN process \cite{Mukhanov,Andreas,Kofman,Shtanov,Kofman1,Bassett}. First idea of the reheating was coined by Linde \cite{Linde}. After the end of inflation the inflaton field oscillates to the minima of the potential and decays into the elementary particles \cite{Rehagen}. These particles interact with each other through some suitable coupling and reheat the universe and achieves a temperature $T_{re}$ which is the reheating temperature and this process is called reheating. The process of reheating can be studied either by perturbative reheating or through parametric resonance called preheating \cite{Lozanov,Martin}. In this article, rather than focussing on the actual mechanism behind the reheating for the power law plateau model, we tried to understand epoch indirectly.
If one consider equation of state ($\omega_{re}$) to be constant during the reheating epoch, the relation between the energy density and the scale factor using  $\rho \propto a^{-3(1+w)}$, can be established as   \cite{deFreitas,Cook,Kai,Gong}
\begin{equation}
    \frac{\rho_{end}}{\rho_{re}} = \left(\frac{a_{end}}{a_{re}} \right)^{-3(1+w_{re})}.
    \label{re1}
\end{equation}
Here subscript $end$ is defined as the end of inflation and $re$ is the end of reheating epoch. Replacing $\rho_{end}$ by $(7/6) V_{end}$ following \cite{Bhattacharya} one can write
\begin{equation}
N_{re} = \frac{1}{3(1+w_{re})} \ln \left(\frac{\rho_{end}}{\rho_{re}} \right)= \frac{1}{3(1+w_{re})} \ln \left(\frac{7}{6}\frac{V_{end}}{\rho_{re}} \right),
    \label{re2}
\end{equation}
We follow the  standard relation between density and temperature:   
\begin{equation}
\rho_{re} = \frac{\pi^2}{30} g_{re} T_{re}^4.
\label{re3}
\end{equation}
Here $g_{re}$ is the number of relativistic species at the end of reheating.
Using (\ref{re1}) and (\ref{re2}), we can establish the relation between $T_{re}$ and $N_{re}$ as:
\begin{equation}
N_{re} = \frac{1}{3(1+w_{re})} \ln \left(\frac{35  V_{end}}{\pi^2 g_{re} T_{re}^4 } \right)
\label{re4}
\end{equation}
Assuming that the entropy is conserved between the reheating  and today, we can write  
\begin{equation}
T_{re}= T_0 \left(\frac{a_0}{a_{re}} \right) \left(\frac{43}{11 g_{re}} \right)^{\frac{1}{3}}=T_0 \left(\frac{a_0}{a_{eq}} \right) e^{N_{RD}} \left(\frac{43}{11 g_{re}} \right)^{\frac{1}{3}},
\label{re5}
\end{equation}

Here $N_{RD}$ is the number of e-folds during radiation era and $e^{-N_{RD}}\equiv a_{re}/a_{eq}$. The ratio $a_{0}/a_{eq}$ can be written as 
\begin{equation}
\frac{a_0}{a_{eq}} = \frac{a_0 H_{k}}{k} e^{-N_{k}} e^{- N_{re}} e^{- N_{RD}}\
\label{re6}
\end{equation}
Using the relation $k_{}=a_{k} H_{k}$ and using the Eq.~(\ref{re4}), (\ref{re5}) and (\ref{re6}), assuming $w_{re} \neq \frac{1}{3}$ and $g_{re}\approx 226$ (degrees of freedom for a supersymmetric model), we can establish the expression for $N_{re}$
\begin{equation}
N_{re}= \frac{4}{ (1-3w_{re} )}   \left[61.5475  - \ln \left(\frac{ V_{end}^{\frac{1}{4}}}{ H_{k} } \right)  - N_{k}   \right]
\label{re7}
\end{equation}
Here we have used Planck's pivot  ($k$) of order $0.02 \; \mbox{Mpc}^{-1}$. In a similar way we can calculate $T_{re}$:
\begin{equation}
T_{re}= \left[ \left(\frac{43}{11 g_{re}} \right)^{\frac{1}{3}}    \frac{a_0 T_0}{k_{}} H_{k} e^{- N_{k}} \left[\frac{35  V_{end}}{\pi^2 g_{re}} \right]^{- \frac{1}{3(1 + w_{re})}}  \right]^{\frac{3(1+ w_{re})}{3 w_{re} -1}}.
\label{re8}
\end{equation}
To evaluate $N_{re}$ and $T_{re}$ first we need to calculate the $H_{k}$, $N_{k}$ and $V_{end}$ for the given potential. Now $H_{k}$ can be represented in terms of $r$ and $A_s$ as:
\begin{equation}
{H_k}= \left[{\frac{1}{6} \sqrt{\frac{\rho }{3}} \left(\pi ^2 {P_s} {r}\right)}\right]^{\frac{1}{3}} 
\label{Hk}
\end{equation}

We restrict our reheating analysis for the class of potentials which satisfies the Trans Planckian Censorship Conjecture \cite{Bedroya,Bedroya1} like potentials for the choice of $n=3,q=1$ and  $n=3,q=2$ . We have plotted $N_{re}$ and $T_{re}$ as a function of $n_{s}$, for the mentioned potential it is not possible to write $r$ as a function of $n_{s}$. So here we will use the numerical method to establish the relation between $r$ and $n_{s}$. For a range of e-folding and doing the necessary cubic fitting we can write the two equations as:

(\textbf{For $n$ = 3 and $q$ = 1 })

\begin{equation}
r = -1.62561\times{10^{-13}} N_e^3 + 4.31547\times{10^{-11}} N_e^2 - 4.02308\times{10^{-9}} N_e + 1.43247\times{10^{-7}} 
\label{rr31}
\end{equation}
\begin{equation}
n_{s} = 9.71953 \times{10^{-8}} N_e^3 - 0.0000269363 N_e^2 + 0.0027286 N_e + 0.879288
\label{nss31}
\end{equation}
  
(\textbf{For $n$ = 3 and $q$ = 2 } ) 

\begin{equation}
r = -2.14466\times{10^{-13}} N_e^3 + 5.69341\times{10^{-11}} N_e^2 - 5.30772\times{10^{-9}} N_e + 1.87807\times{10^{-7}} 
\label{rr32}
\end{equation}
\begin{equation}
n_{s} = 9.91911 \times{10^{-8}} N_e^3 - 0.0000274979 N_e^2 + 0.00278734 N_e + 0.876415
\label{nss32}
\end{equation}
    
\begin{figure}[ht!]
  \centering
\begin{minipage}[b]{0.4\textwidth}
\includegraphics[width=2.9in, height=2.7in] {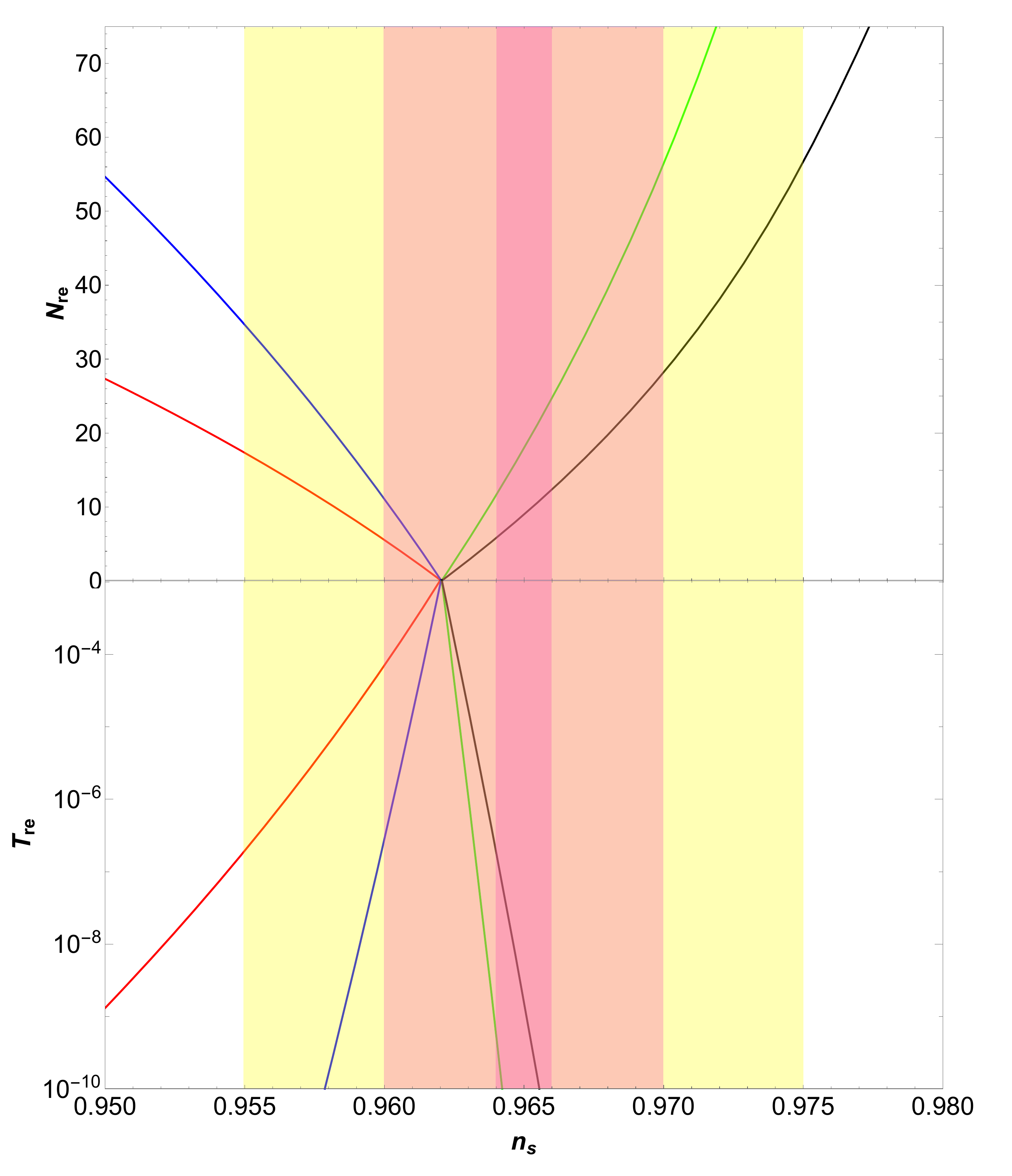}
\caption{For the choice of  $n=3,q=1$, Plots of $N_{re}$ and $T_{re}$ as a function $n_s$ for different values of $w_{re}$. The red line corresponds to $w_{re}= -1/3$, the blue line corresponds to $w_{re}= 0$, the green line corresponds to $w_{re}= 2/3$ and finally the black line corresponds to $w_{re}= 1$. The light pink shaded region corresponds to the 1-$\sigma$ bounds and yellow shaded region corresponds to the 2-$\sigma$ bounds  on $n_s$ from {\it Planck'18} [TT,TE, EE+lowE+lensing+BK15]. The dark pink shaded region corresponds to the 1-$\sigma$ bound of future CMB observations \cite{euclid, prism} using same central value for $n_s$.} 
\label{Tre_Nre31}
\end{minipage}
\hfill
\begin{minipage}[b]{0.4\textwidth}
  \includegraphics[width=2.9in, height=2.85in]{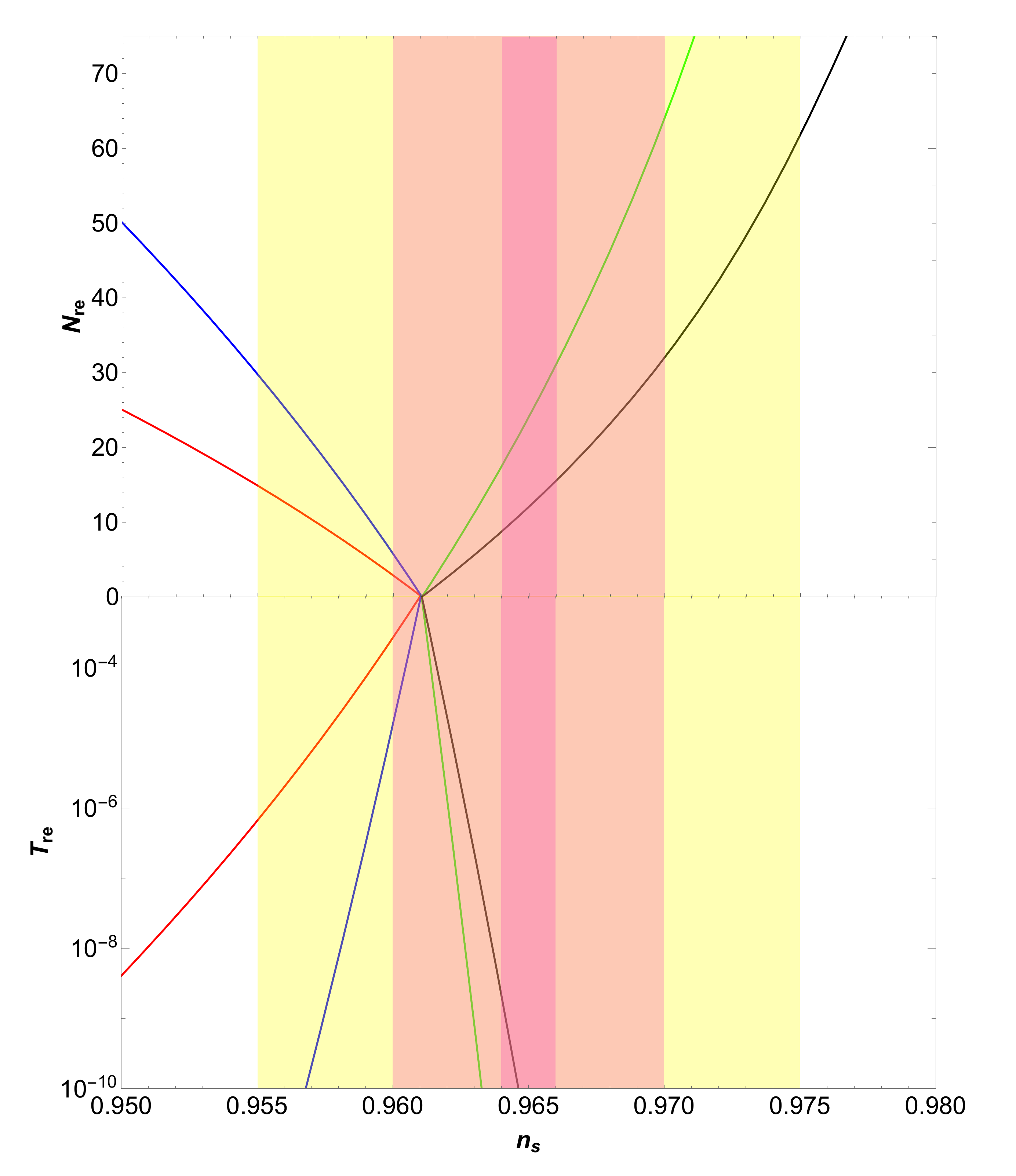}
 \caption{For the choice of  $n=3,q=2$, Plots of $N_{re}$ and $T_{re}$ as a function $n_s$ for different values of $w_{re}$. The red line corresponds to $w_{re}= -1/3$, the blue line corresponds to $w_{re}= 0$, the green line corresponds to $w_{re}= 2/3$ and finally the black line corresponds to $w_{re}= 1$. The light pink shaded region corresponds to the 1-$\sigma$ bounds and yellow shaded region corresponds to the 2-$\sigma$ bounds  on $n_s$ from {\it Planck'18} [TT,TE, EE+lowE+lensing+BK15]. The dark pink shaded region corresponds to the 1-$\sigma$ bound of future CMB observations \cite{euclid, prism} using same central value for $n_s$}
 \label{Tre_Nre32}
\end{minipage}
\end{figure}
Using the above Eq.(\ref{rr31} - \ref{nss32}), we can write the Eq. (\ref{Hk}) as a function of $n_s$.  From the end of inflation condition $\epsilon_{RS} = 1$ , one can calculate the $V_{end}$ and then get $T_{re}$ and $N_{re}$  by using Eq.~(\ref{re7}), (\ref{re8}) and (\ref{Hk}) for different values of equation of state ($w_{re}$). The changes of $T_{re}$ and $N_{re}$ for different values of $w_{re}$ is shown in fig. (\ref{Tre_Nre31},\ref{Tre_Nre32}). We would like to mention that the merging points for the $T_{re}$ plot and the $N_{re}$ plot correspond to the instant reheating scenario thus making $N_{re}=0$
\section{Bound on tensor to scalar ratio from TCC }
\setcounter{equation}{0}
\setcounter{figure}{0}
\setcounter{table}{0}
\setcounter{footnote}{1}
\label{tcccal}
\begin{figure}
    \centering
    \subfigure[]{\includegraphics[width=0.4\textwidth]{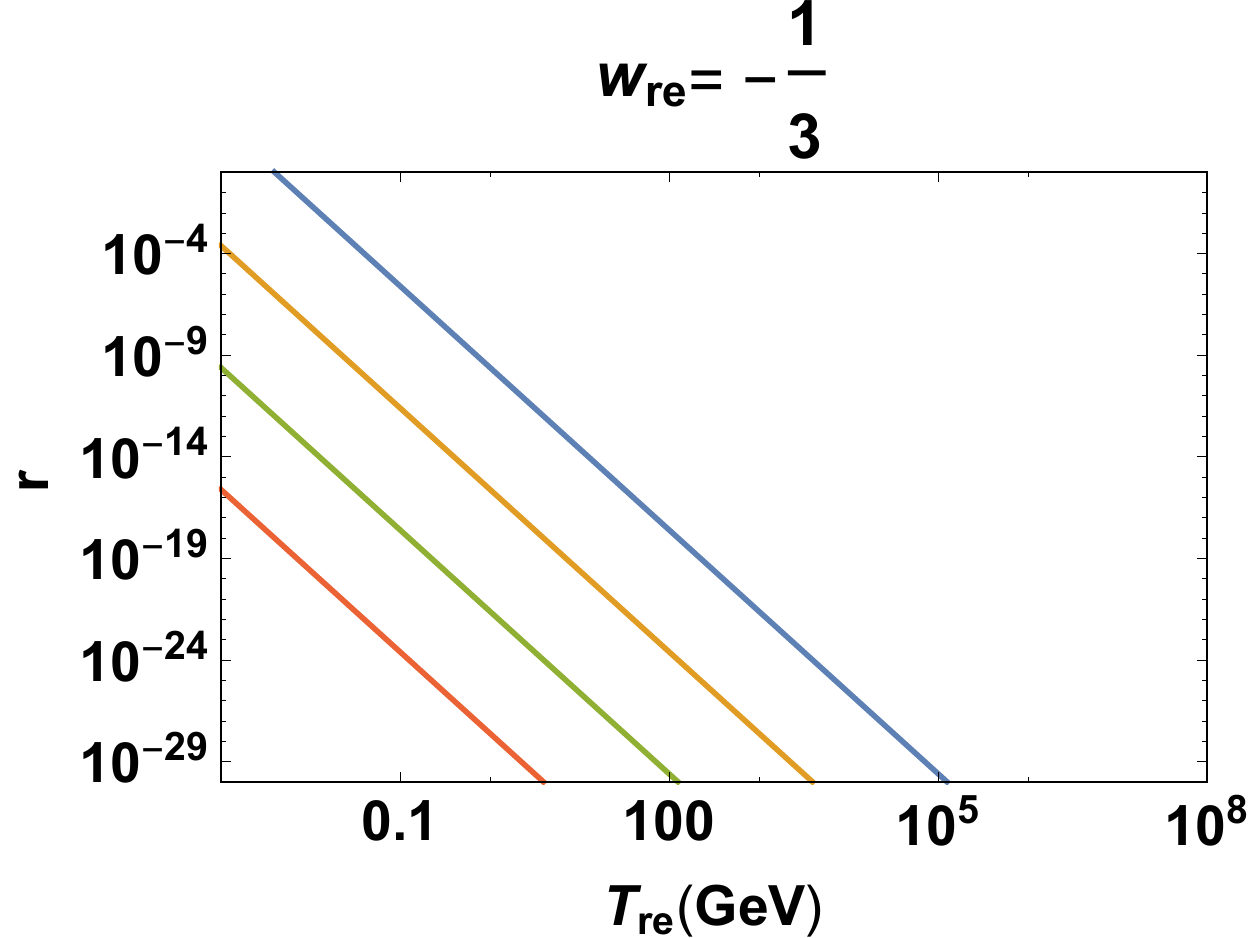}} \label{fig14}
    \subfigure[]{\includegraphics[width=0.4\textwidth]{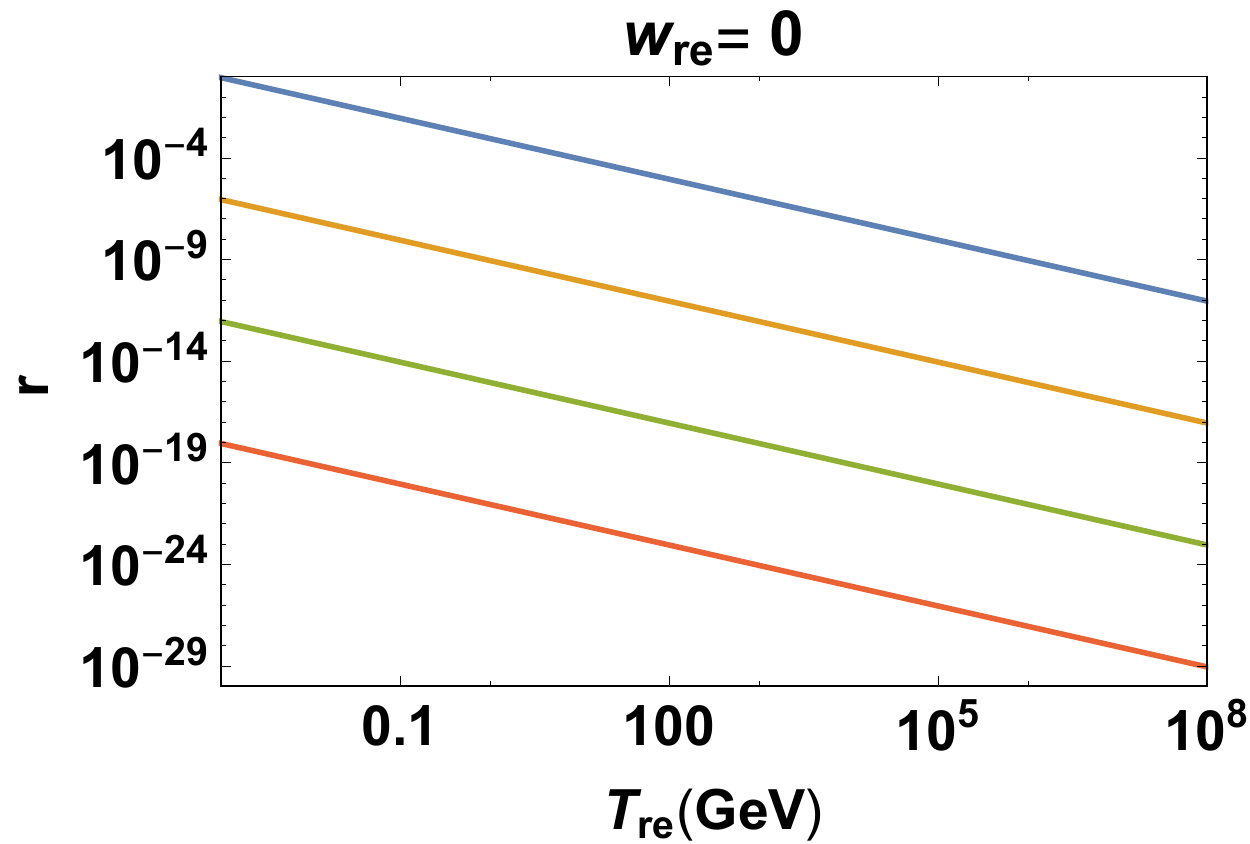}} \label{fig24}
    \subfigure[]{\includegraphics[width=0.4\textwidth]{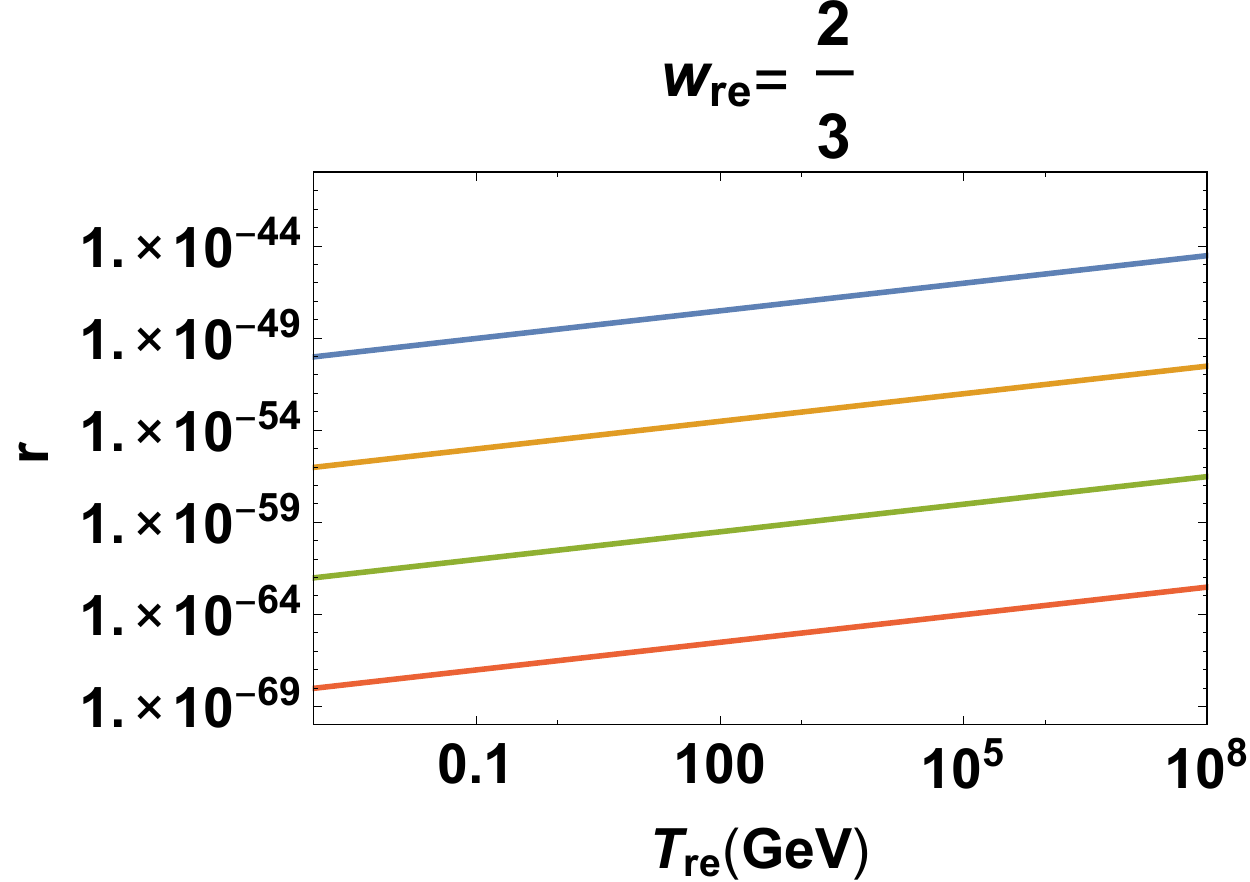}}\label{fig34}
    \subfigure[]{\includegraphics[width=0.4\textwidth]{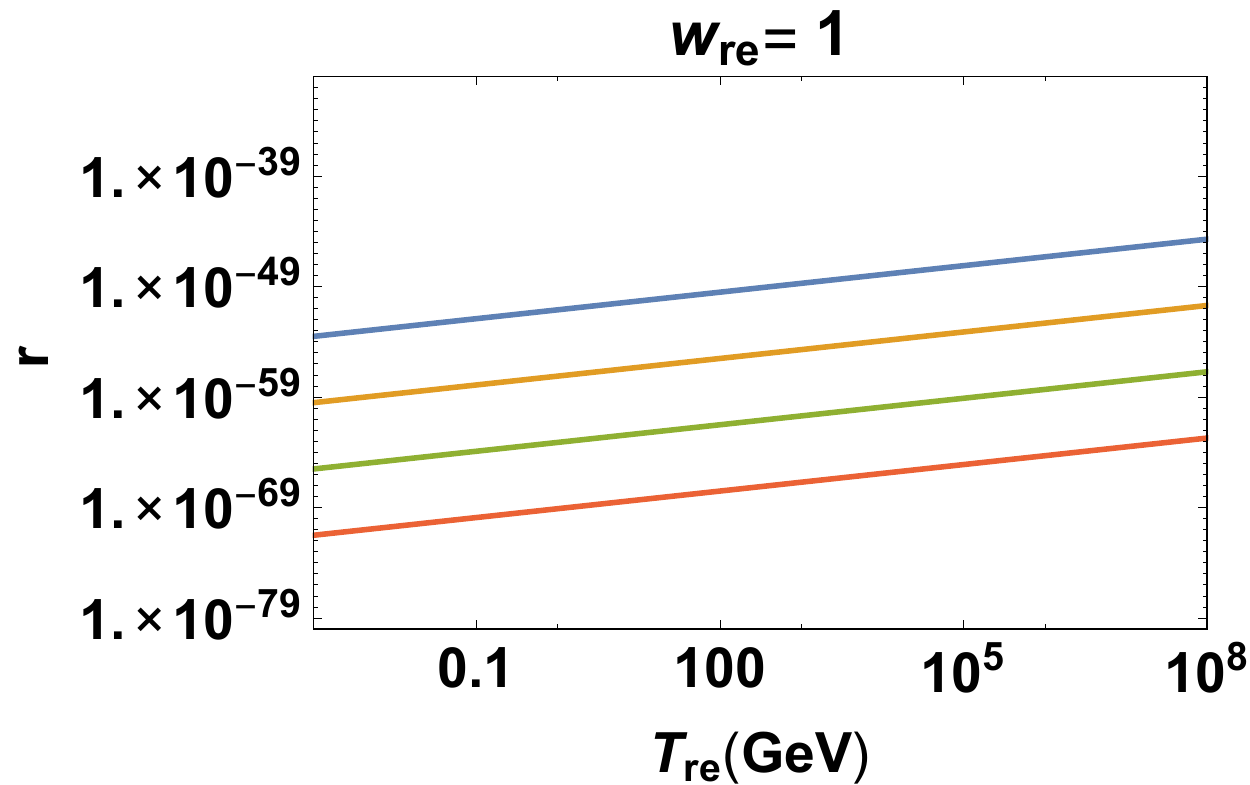}}\label{fig44}
    \caption{Plots of $r$ versus $T_{re}$ for some given values of $M_5$ allowed in the range following \cite{Bhattacharya} with $w_{re}$ fixed in each case. The red, green, yellow and blue lines corresponds to $M_5= 10^{17},10^{16},10^{15},10^{14} GeV$ respectively.}
    \label{figtcc}
\end{figure} 

From  \cite{Mizuno}, for a general value of the $w_{re}$, one can write the relation between $H_k$ and $T_{re}$ as:
\begin{equation}
ln \frac{H_k}{M_p}< \frac{3 (1+ w_{re})}{2(2+ 3w_{re})}ln \frac{H_0}{T_0}+ \frac{1- 3w}{2(2+ 3w)}ln \frac{M_p}{T_{reh}},
\end{equation}
where $H_0$ is the Hubble constant and $T_0$ is the current CMB temperature. Then for different $w_{re}$, one gets:
\begin{equation}
 \frac{H_k}{M_p}< \frac{1.50072\times 10^{-11}}{T_{re}},~~~~for~~w_{re}= -\frac{1}{3}~,
\end{equation}
\begin{equation}
 \frac{H_k}{M_p}< \frac{4.85615\times 10^{-18}}{T_{re}^{1/4}},~~~~for,~~w_{re}= 0~,
\end{equation}
\begin{equation}
 \frac{H_k}{M_p}< Exp[\frac{1}{8}[-336.326-Ln(\frac{2.42\times 10^{18}}{T_{re}})]],~~~~for,~~w_{re}= \frac{2}{3}~,
\end{equation}
\begin{equation}
 \frac{H_k}{M_p}< Exp[\frac{1}{5}[-201.795-Ln(\frac{2.42\times 10^{18}}{T_{re}})]],~~~~for,~~w_{re}= 1~,
\end{equation}
where all the inequalities are calculated for $H_0= 68 (km/s)/Mpc$ and $T_0= 2.73K$ after converting to the $GeV$ unit.\\
Now for RS $II$ braneworld the tensor to scalar ratio can be calculated using Eq.~(\ref{Ps} -\ref{fx}) as:
\begin{equation}
r= \frac{P_T}{P_s}=\frac{8\bigg(\frac{H}{2 \pi}\bigg)^2 F(x_0)^2}{P_s(k_{pivot})}~,
\label{rtre}
\end{equation}
where, $P_S(k_{pivot})$ is fixed at the pivot scale $k_{pivot}= 0.02 Mpc^{-1}$  \cite{Aghanim}. Obviously the running and the following corrections to the curvature perturbation is neglected without loosing too much generality. From Eq.~(\ref{rtre}) and Eq.~(\ref{rho}) one gets the relation between $r$ and $T_{re}$ for a fixed value of $M_5$. The relationship for different $w_{re}$ is depicted in the fig.~(\ref{figtcc}). From (\ref{figtcc}) it can be observed that for $w_{re}= -1/3$ and $w_{re}=0$ for low values of $T_{re}$ (but greater than the $T_{BBN}\cong 10 MeV$) with lowest value of $M_5$ allowed from \cite{Bhattacharya}, one can get $r$ in the range of $10^{-4}$ to $10^{-3}$. Thus, one can see from fig.~(\ref{Tre_Nre31},\ref{Tre_Nre32}), at least for $w_{re}= 0$, we can get high value of $r$ corresponding to $T_{re}$ consistent with the 2-$\sigma$ bound on $n_s$ from {\it Planck'18 } [TT,TE, EE+lowE+lensing+BK15].\\
Thus, in a nutshell in this paper we have shown the effect of TCC in case of RS $II$ braneworld inflationary model following \cite{Mizuno} which was done for the standard case. The interesting conclusion in RS $II$ braneworld is that one can actually increase the upper bound on $r$. For the particularly chosen model of inflation, in this work, one can see for a correct choice of $w_{re}$ one can circumvent the so called TCC crisis.



\section{Conclusion} 
\label{sec:conl}

In this article, we have studied the power law plateau model in the light of recent CMB observations. We have shown that in the RS $II$ brane world scenario it is easy to evade the Swampland Conjuncture and for certain choices of inflationary parameters we can also satisfy the Trans Planckian Censorship Conjecture. For the choice of $n=1,q=1$ to $n=4,q=4$ all the cosmological observables are well within $Planck'18$ [TT,TE, EE+lowE+lensing+BK15] bounds and Swampland conjecture can be satisfied. We have taken the number of e-foldings between $55-75$, which are consistent with the RS $II$ braneworld \cite{Wang}.  For  certain choices of $n$ and $q$ it is possible to satisfy the TCC which is shown in the Table (2.1), with tensor to scalar  ratio$(r)$ of the order of $10^{-8}$. We have also studied the reheating epoch. Though the reheating analysis can be done for all the potentials but we have shown for only two cases which also satisfies the TCC. In case of the power law plateau  model of inflation, consistency with SC makes $M_5$ of the order of $10^{-2}M_P$ which is consistent with \cite{Bhattacharya}. Though we have shown that the class of potentials is consistent with observation while satisfying swampland, it will be interesting to get the correct parameter estimation of the class of this model directly from observation. We would like to come back to this in future.

\begin{acknowledgments}
 Work of MRG is supported by Department of Science and Technology, Government of India under the Grant Agreement number IF18-PH-228 (INSPIRE Faculty Award). The authors would like to thank Mostafizur Rahman and Nur Jaman for useful discussion.
 \end{acknowledgments}



\end{document}